\documentclass[12pt,english,floatfix,showkeys,superscriptaddress,aps,prd,preprint]{revtex4}
\usepackage[latin1]{inputenc}
\usepackage[T1]{fontenc}
\usepackage{lmodern}
\usepackage{amsmath}
\usepackage{amssymb}
\usepackage{graphicx}
\usepackage{float}
\usepackage{esint}
\usepackage{longtable}
\usepackage{dcolumn}
\usepackage{babel}
\usepackage{csquotes}
\usepackage{color}
\usepackage{slashed}
\usepackage{simplewick}
\usepackage{amsmath,latexsym}

\usepackage{hyperref}
\hypersetup{
    colorlinks,
    citecolor=blue,
    filecolor=green,
    linkcolor=purple,
    urlcolor=red,
}

\usepackage{slashed}

\usepackage{hyperref}
\hypersetup{colorlinks,breaklinks,
			citecolor=[rgb]{0,0.0,1.0},
            urlcolor=[rgb]{0.0,0.0,1.0},
            linkcolor=[rgb]{0,0.5,0.9}}

\begin{document}

\title{Yang-Mills Casimir wormholes in $D=2+1$}

\author{Alana C. L. Santos}
\email{alanasantos@fisica.ufc.br}
\affiliation{Universidade Federal do Cear\'a (UFC), Departamento de F\'isica,\\ Campus do Pici, Fortaleza - CE, C.P. 6030, 60455-760 - Brazil.}
\author{C\'{e}lio R. Muniz}
\email{celio.muniz@uece.br}
\affiliation{Universidade Estadual do Cear\'a (UECE), Faculdade de Educa\c{c}\~ao, Ci\^encias e Letras de Iguatu, Av. D\'ario Rabelo s/n, Iguatu-CE, 63.500-00 - Brazil.}
\author{Roberto V. Maluf}
\email{r.v.maluf@fisica.ufc.br}
\affiliation{Universidade Federal do Cear\'a (UFC), Departamento de F\'isica,\\ Campus do Pici, Fortaleza - CE, C.P. 6030, 60455-760 - Brazil.}
\affiliation{Departamento de F\'{i}sica Te\'{o}rica and IFIC, Centro Mixto Universidad de Valencia - CSIC. Universidad
de Valencia, Burjassot-46100, Valencia, Spain.}



\date{\today}

\begin{abstract}
This work presents new three-dimensional traversable wormhole solutions sourced by the Casimir density and pressures related to the quantum vacuum fluctuations in Yang-Mills (Y-M) theory. We begin by analyzing the noninteracting Y-M Casimir wormholes, initially considering an arbitrary state parameter $\omega$ and determine a simple constant wormhole shape function. Next, we introduce a new methodology for deforming the state parameter to find well-behaved redshift functions. The wormhole can be interpreted as a legitimate Casimir wormhole with an expected average state parameter of $\omega=2$. Then, we investigate the wormhole curvature properties, energy conditions, and stability. Furthermore, we discover a novel family of traversable wormhole solutions sourced by the quantum vacuum fluctuations of interacting Yang-Mills fields with a more complex shape function. Deforming the effective state parameter similarly, we obtain well-behaved redshift functions and traversable wormhole solutions. Finally, we examine the energy conditions and stability of solutions in the interacting scenario and compare to the noninteracting case.
\end{abstract}


\maketitle


\section{Introduction}

The Casimir effect is associated with quantum vacuum fluctuations when we impose boundary conditions on quantum fields. The first configuration analyzed, two neutral parallel conducting plates separated by a very short distance, produces an attractive force \cite{Casimir:1948dh}. However, the nature of this force is strongly dependent on the geometry and dimension in question and can be repulsive, for example, \cite{Boyer:1968uf}. This manifestation of quantum effects on macroscopic scales was confirmed in 1958 \cite{Sparnaay:1958wg} and has been subjected to increasingly precise experiments \cite{C1, C2, C3}. Since then, this effect has been intensely studied in the literature from different perspectives \cite{Santos:2021jjs, Muniz:2015jba, Lima:2019pbo, Maluf:2019ujv}, including from generalizations of classical electromagnetism \cite{Garattini:2021kca,deOliveira:2022hew}. In this sense, we highlight the analyses performed in the context of a non-Abelian gauge theory in $2+1$ dimensions via Hamiltonian formulation \cite{Karabali:2018ael}, and lattice simulations based on first principles \cite{Chernodub:2018pmt} in which the authors identify vacuum backreaction effects on the plates.

On the other hand, a distinct characteristic of the Casimir energy density is the violation of energy conditions due to its negative value for some configurations. This fact promotes it as one of the scarce energy sources capable of sustaining traversable wormholes \cite{Tripathy:2020ehi} - solutions of Einstein's equations analogous to a hypothetical tunnel in spacetime between two distant points in the universe, whose traversability requires exotic matter \cite{Morris:1988cz} (except for some modified theories \cite{Pavlovic:2014gba},  which has been useful for detecting these structures \cite{DeFalco:2021klh,DeFalco:2021ksd}). Interestingly, the pioneering solution in $3+1$ dimensions was only found in 2019, with the careful work of Garattini \cite{Garattini:2019ivd} and, in the sequence, extended to $d$ dimensions \cite{Oliveira:2021ypz}. However, Alencar, Bezerra, and Muniz showed that in $2+1$ dimensions, the Casimir energy density and pressure are not capable of sustaining this structure. Therefore, we would not have ``pure'' Casimir wormholes in this reduced dimension \cite{Alencar:2021ejd}.

Working with gravity models in lower dimensions makes sense since the phenomenon of spontaneous dimensional reduction of spacetime has garnered considerable attention in recent years. It refers to the intriguing notion that the number of dimensions in our familiar four-dimensional spacetime may decrease at extremely high energy scales. This concept challenges our conventional understanding of the fabric of the Universe, suggesting that at extreme conditions, the fundamental structure of spacetime may undergo a remarkable transformation \cite{Carlip:2016qrb}.

In this sense, inspired by works in which quarkonic matter is a source for wormholes \cite{Lobo}, we will seek to approach this question again, but considering a gluonic scenario, through the potentials found for short and long distances between perfect chromoelectric conductors in the Yang-Mills theory mentioned before. As we will see, we again find a non-finite redshift function throughout space, which compromises one of the traversability conditions established by Morris and Thorne \cite{Morris:1988cz}. Some methods adopted in the literature are associated with modifications in the metric \cite{Garattini:2019ivd} or in the Casimir radial density/pressure \cite{Alencar:2021ejd}. In this work, we propose a new method that makes a slight distortion in the equation of state by adding a function obeying an inverse power law of the radial coordinate. By averaging our deformation, we identify that the original equation of state is maintained, and therefore, we would have a legitimate Casimir source on average. We will use this approach for noninteracting and interacting (confined) Yang-Mills fields.  It is worth mentioning that Chew and Lim have recently obtained numerical solutions for wormholes using the phantom field to support the throat in the Bogomol'nyi-Prasad-Sommerfield limit and beyond, for the Einstein-Yang-Mills-Higgs action \cite{Chew:2021vxh,Chew:2020svi}.

Our paper is organized as follows: In Sec. \ref{sec2}, we find a family of traversable wormhole solutions considering the usual form of Casimir energy density and our deformation method. Besides, we identified the characteristics of the matter sources that produce these solutions. In Sec. \ref{sec3}, we use our deformation method again to find new wormhole solutions with Casimir energy for the $SU(N)$ gauge theory, and we investigate the properties of the geometry and matter-energy content in these solutions. Finally, we outline our perspectives and conclusions in Sec. \ref{conclusion}.

\section{Noninteracting Case\label{sec2}}

In this section, we will revisit the issue of whether it is feasible
to maintain a traversable wormhole in a $(2+1)D$ spacetime
using the Casimir quantities - energy density and pressure - as observed
in the $(3+1)D$ scenario \cite{Garattini:2019ivd}. This question was first addressed
by Alencar et al. in Ref. \cite{Alencar:2021ejd} in the context of a
Casimir wormhole embedding in a graphene sheet. As we will
see below, the answer to this question is negative, at least if we
consider the usual $(2+1)D$ Casimir energy and pressure. To overcome
this limitation, we will present a new method for generating traversable
wormholes in this scenario. 

Our starting point is the Einstein field equations in 2+1 dimensions:
\begin{equation}
G_{\mu\nu}=\kappa T_{\mu\nu},
\end{equation}
where $\kappa=8\pi\tilde{G}/c^{4}$ with $\tilde{G}$ being the Newtonian gravitational constant in two spatial dimensions. In the natural system of units, $\tilde{G}$ has units of length $[\tilde{G}]=\ell$. For simplicity, we consider this arbitrary constant to have a magnitude equal to 1. In addition, $T_{\mu\nu}$ represents the energy-momentum for the matter source. 

Let us assume a circularly symmetric and static ansatz for the spacetime
metric with the line element given by
\begin{equation}
ds^{2}=-e^{2\Phi(r)}dt^{2}+\frac{1}{1-\frac{b(r)}{r}}dr^{2}+r^{2}d\theta^{2},\label{eq:metricAnsatz}
\end{equation}
describing a (2+1)-dimensional Morris-Thorne wormhole in which the
redshift function $\Phi(r)$ and the shape function $b(r)$ are arbitrary
functions of the polar coordinate $r\in\left[r_{0},+\infty\right)$ \cite{Perry:1991qq}. Thus, the coordinate $r$
must be decreases from infinity to a minimum value $r_{0}$ , the
radius of the throat. These functions must satisfy some properties
to ensure a traversable wormhole solution: $(i)$ a flaring-out condition,
determined by the minimality of the wormhole throat, such that $(b-b'r)/b^{2}>0$,
and at the throat, $b(r_{0})=r_{0}$; $(ii)$ the condition $1-b/r\geq0$
must also satisfy to guarantee wormhole solutions; $(iii)$ it must
be ensured that there are no horizons present, which are identified
as surfaces with $e^{\Phi}\rightarrow0$, so that $\Phi(r)$ is finite
everywhere \cite{Morris:1988cz}. 

The Einstein field equations under the metric ansatz (\ref{eq:metricAnsatz})
have the simple forms:
\begin{align}
G_{\ t}^{t}= & \frac{rb'-b}{2r^{3}}=\kappa\rho(r),\label{eq:g00}\\
G_{\ r}^{r}= & \frac{\left(r-b\right)\Phi'}{r^{2}}=\kappa p_{r}(r),\label{eq:grr}\\
G_{\ \theta}^{\theta}= & \left(1-\frac{b}{r}\right)\left[\Phi''+(\Phi')^{2}\right]+\frac{\left(b-rb'\right)\Phi'}{2r^{2}}=\kappa p_{\theta}(r),\label{eq:gthetatheta}
\end{align}
where $T_{\ \nu}^{\mu}=\mbox{diag (-\ensuremath{\rho,p_{r},p_{\theta}})}$
in which $\rho(r)$ is the surface energy density, $p_{r}(r)$ is
the radial pressure, and $p_{\theta}(r)$ is the lateral pressure.
Additionally, the covariant energy-momentum conservation law leads to 
\begin{equation}
p_{r}^{'}=\frac{1}{r}(p_{\theta}-p_{r})-(\rho+p_{r})\Phi^{'}.\label{eq:conservation}
\end{equation}

From Eq. (\ref{eq:g00}), we note that the flare-out condition $(i)$
is verified if $\rho(r)<0$. Remembering that the Casimir effect is
typically produced by a negative energy density, we could argue that
Casimir energy is a natural exotic matter source for traversable wormholes. Hence, we will assume that our matter source is due to the Casimir energy density, following the ideas of Garattini in Ref. \cite{Garattini:2019ivd} closely. 

Usually, the Casimir energy density in $(2+1)$-spacetime dimensions can be represented
in the form
\begin{equation}
\rho(r)=-\frac{\lambda}{r^{3}},\label{eq:density}
\end{equation}
where $\lambda$ is a positive constant which depend on the specific
model considered. For instance, an abelian gauge field confined between
two parallel static wires exhibits a Casimir parameter $\lambda=\zeta(3)/16\pi$,
while the same geometry applied to the non-Abelian $SU(N)$ gauge
field leads to $\lambda=(N^{2}-1)\zeta(3)/16\pi$ (noninteracting case),
where $\zeta(x)$ is the zeta function \cite{Chernodub:2018pmt}.

Once the Casimir force is obtained by the differentiation of
the Casimir Energy, it is straightforward to identify the Casimir
radial pressure as \cite{Alencar:2021ejd}
\begin{equation}
p_{r}^{\text{Cas}}(r)=-2\frac{\lambda}{r^{3}},\label{eq:EoS}
\end{equation}
which leads to an equation of state (EoS) given by $p_{r}=\omega\rho$
with $\omega=2$. For our present purpose, let us relax this condition and work with a generic $\omega$, taking the Casimir limit at an opportune time.

From (\ref{eq:g00}) and (\ref{eq:density}), we obtain the following shape function 
\begin{equation}
b(r)=2\kappa\lambda+\left(1-\frac{2\kappa\lambda}{r_{0}}\right)r,\label{eq:bsolution}
\end{equation}
where we take into account the boundary condition $b(r_{0})=r_{0}.$
Substituting (\ref{eq:bsolution}) into (\ref{eq:grr})
and using the EoS $p_{r}=\omega\rho$, we find the redshift function
in the form
\begin{equation}
\Phi(r)=\Phi_{0}+\frac{\omega}{2}\left[\ln r-\ln(r-r_{0})\right],\label{eq:phisolution}
\end{equation}
where $\Phi_{0}$ is an arbitrary integration constant. In order to
obtain the appropriate asymptotic flat limit, it is convenient to
fix $\Phi_{0}=0$ such that $\lim_{r\rightarrow\infty}\Phi(r)=0.$

Employing the shape function (\ref{eq:bsolution}) and the redshift function (\ref{eq:phisolution}), we can write the line element (\ref{eq:metricAnsatz})
in the following way:
\begin{equation}
ds^{2}=-\left(\frac{r}{r-r_{0}}\right)^{\omega}dt^{2}+\frac{1}{\frac{2\lambda\kappa}{r_{0}}\left(1-\frac{r_{0}}{r}\right)}dr^{2}+r^{2}d\theta^{2}.\label{eq:line1}
\end{equation}

Unfortunately, due to the presence of the term $\ln(r-r_{0})$ in
(\ref{eq:phisolution}), the line element (\ref{eq:line1}) is not
expected to represent a traversable wormhole. In fact, for $\omega>0$,
there is no event horizon since $\left(\frac{r}{r-r_{0}}\right)^{\omega}=0$
does not have a simple zero at $r=r_{H}$. Therefore, we have neither
a black hole nor a traversable wormhole, as $\Phi(r)$ is not finite
everywhere and diverges at $r=r_{0}$. Instead, we have a singularity
described by the following line element (see \cite{Garattini:2019ivd}, Appendix A):
\begin{equation}
ds^{2}=-\frac{1}{\left(1-\frac{r_{0}}{r}\right)^{\omega}}dt^{2}+\frac{1}{\frac{2\lambda\kappa}{r_{0}}\left(1-\frac{r_{0}}{r}\right)}dr^{2}+r^{2}d\theta^{2}.
\label{NS}\end{equation}
Furthermore, in the case $\omega<0$, we have one horizon at $r=r_{0}$
($g_{tt}=0$ has a simple zero at $r=r_{0}$), and for the particular
value $\omega=-1$ , we find one black Hole solution given by
\begin{equation}
ds^{2}=-\left(1-\frac{r_{0}}{r}\right)dt^{2}+\frac{1}{\frac{2\lambda\kappa}{r_{0}}\left(1-\frac{r_{0}}{r}\right)}dr^{2}+r^{2}d\theta^{2}.
\end{equation}
As pointed out in Ref. \cite{Garattini:2019ivd}, the presence of a logarithmic term
in the redshift function is also observed in the $(3+1)$-dimensional
case. However, unlike in that case, we cannot eliminate this ``annoying''
term because in $(2+1)$ dimensions, we have less freedom
to fix the free parameters of the system. The metric (\ref{eq:line1})
was also obtained from Ref. \cite{Alencar:2021ejd} with $\omega=-2$, in which
the authors added a cosmological constant to work around the problem
that was pointed out. Here, we propose a different strategy. Let us
assume a distortion of the EoS in the following way:
\begin{equation}
\omega\rightarrow\omega+\epsilon f(r),\label{eq:prescriptionomega}
\end{equation}
where $\epsilon$ is a free parameter that is responsible for deviating
$\omega$ from the original Casimir prescription. The function $f(r)$
can be chosen to obtain the desired wormhole solution.

With prescription (\ref{eq:prescriptionomega}), equation (\ref{eq:grr})
assume the following form
\begin{equation}
\Phi'(r)=-\frac{r_{0}(\omega+\epsilon f(r))}{2r(r-r_{0})}.\label{eq:Phieq}
\end{equation}
A simple and convenient choice for the function $f(r)$ able to accomplish
our aforementioned objectives is $f(r)=-1/r$. Therefore, the redshift
function becomes
\begin{equation}
\Phi_{\epsilon}(r)=\Phi_{0}+\frac{1}{2r_{0}}\left[\frac{\epsilon r_{0}}{r}-(\epsilon-\omega r_{0})\ln(r)+(\epsilon-\omega r_{0})\ln(r-r_{0})\right],
\end{equation}
and, similar to the $(3+1)$-dimensional case \cite{Garattini:2019ivd}, the
term $\ln(r-r_{0})$ can be removed by setting the $\epsilon$-parameter
as $\epsilon=\omega r_{0}$. Then, the line element (\ref{eq:metricAnsatz})
takes the form 
\begin{equation}
ds^{2}=-e^{\frac{\omega r_{0}}{r}}dt^{2}+\frac{1}{1-\frac{r_{0}}{r}}dr^{2}+r^{2}d\theta^{2},\label{eq:solTWn=00003D1}
\end{equation}
where $\Phi_{0}=0$, and we fix $r_{0}=2\lambda\kappa=16\pi\lambda\tilde{G}$ for simplicity since the $b(r)$ function assumes a constant form. Now, our solution (\ref{eq:solTWn=00003D1}) satisfies the conditions $(i)-(iii)$ and represents a traversable wormhole with a throat radius on the order of $r_{0}\sim\tilde{G}$ (in natural units).

It is noteworthy that the above result
can be generalized to give rise to a class of Casimir-like traversable
wormholes. This is achieved by taking $f(r)=-1/r^{n}$, where $n$
is a positive integer. Solving Eq. (\ref{eq:Phieq}) for different
values of $n$, we find a family of redshift functions $\Phi_{\epsilon}^{(n)}(r)$
expressed as 
\begin{equation}
\Phi_{\epsilon}^{(2)}(r)=\Phi_{0}+\frac{1}{4r_{0}^{2}}\left[\frac{\epsilon r_{0}}{r^{2}}(2r+r_{0})-2(\epsilon-\omega r_{0}^{2})\ln(r)+2(\epsilon-\omega r_{0}^{2})\ln(r-r_{0})\right],
\end{equation}
\begin{equation}
\Phi_{\epsilon}^{(3)}(r)=\Phi_{0}+\frac{1}{12r_{0}^{3}}\left[\frac{\epsilon r_{0}}{r^{3}}(6r^{2}+3rr_{0}+2r_{0}^{2})-6(\epsilon-\omega r_{0}^{3})\ln(r)+6(\epsilon-\omega r_{0}^{3})\ln(r-r_{0})\right],
\end{equation}
or in the general form
\begin{eqnarray}
&&\Phi_{\epsilon}^{(n)}(r)=\nonumber\\
&&\Phi_{0}+\frac{1}{2a_{1}r_{0}^{n}}\left[\frac{\epsilon r_{0}}{r^{n}}\sum_{p=1}^{n}a_{p}r^{n-p}r_{0}^{p-1} -a_{1}(\epsilon-\omega r_{0}^{n})\ln(r)+a_{1}(\epsilon-\omega r_{0}^{n})\ln(r-r_{0})\right],\label{eq:Phi_n}
\end{eqnarray}where the coefficients $a_{p}(n)$ are determined to satisfy the associated
differential equation (\ref{eq:Phieq}). Note that the $\ln(r-r_{0})$
term is removed in each case by setting $\epsilon=\omega r_{0}^{n}$,
resulting in the general wormhole solution given by 
\begin{equation}
ds^{2}=-\exp\left(\frac{\omega r_{0}}{a_{1}r^{n}}\sum_{p=1}^{n}a_{p}r^{n-p}r_{0}^{p-1}\right)dt^{2}+\frac{1}{1-\frac{r_{0}}{r}}dr^{2}+r^{2}d\theta^{2},\label{eq:lineCasimirAbelian}
\end{equation}
with $n$ being a positive integer. Evidently, our distortion procedure
of the EoS (\ref{eq:EoS}) only affects the redshift function, keeping
the shape function solution (\ref{eq:bsolution}) unaltered. 

It is worth commenting that the promotion of the EoS parameter to be a function of the radial coordinate is a common approach in cosmology in the attempt to model dark energy \cite{Sharov:2022hxi,Barboza:2009ks}. Also, this approach usually happens in the context of a Brans-Dicke type scalar-tensor theories, where the EoS parameter $\omega$ depends on the scalar field profile and hence on the radial coordinate \cite{Fujii:2003pa, Singh:1987is,Lee:2010tm}. Notably, this framework also led to the discovery of wormhole solutions \cite{Franciolini:2018aad,Agnese:1995kd,Lobo:2010sb} . However, our main goal with our study is to find a simple method to create traversable wormholes in $D=2+1$ using the Casimir energy as source matter. To this end, we have proposed to deform the EoS parameter $\omega$ to encompass curvature effects arising from the backreaction between gravity and the matter source to achieve a physically acceptable wormhole solution. Note that asymptotically, the effective EoS parameter follows the usual Casimir relation (\ref{eq:EoS}), while for short distances, it undergoes a correction that enables the formation of traversable wormholes. In this way, the issue of the physical origins of our effective EoS parameter is of interest but lies beyond our present scope.

\begin{figure}[!h]
\begin{center}
\includegraphics[height=7.5cm]{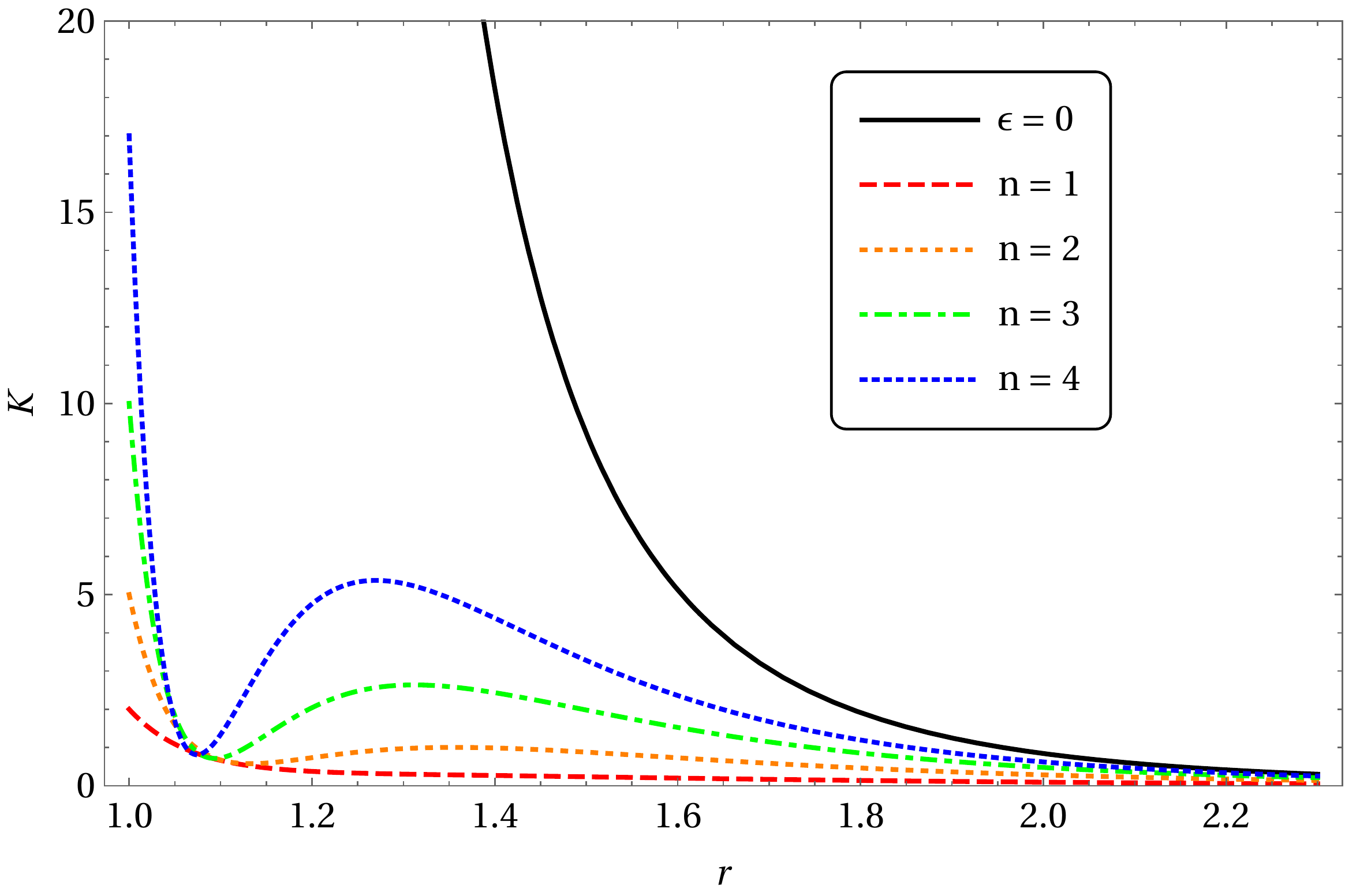}
\end{center}
\caption{Kretschmann scalar associated with the metrics (\ref{NS}) ($\epsilon=0$) and (\ref{eq:lineCasimirAbelian}) for some values of $n$, with $\omega=2$ and $r_{0}=1$. 
\label{fig1}}
\end{figure}

To finish our analysis of the Casimir wormhole solution associated
to the line element (\ref{eq:lineCasimirAbelian}), let us exhibit 
the expression for the Kretschmann scalar $K=R_{\mu\nu\alpha\beta}R^{\mu\nu\alpha\beta}$
for $n=1$ and $\omega=2$, namely, 
\begin{equation}
K=\frac{r_{0}^{2}}{r^{10}}\left(4r_{0}^{4}+12r_{0}^{3}r-3r_{0}^{2}r^{2}-32r_{0}r^{3}+21r^{4}\right).
\end{equation}
As expected, the Kretschmann scalar does not diverge on the interval
$r\in\left[r_{0},+\infty\right)$, and approaches zero at the asymptotic
limit. Furthermore, on the throat, it assumes the constant value of
$\left.K\right|_{r=r_{0}}=\frac{2}{r_{0}^{4}}_{.}$. In Fig. \ref{fig1}, we plot the Kretschmann scalar as a function of the radial coordinate. As expected, when we turn off our correction of the EoS parameter ($\epsilon = 0$), the Kretschmann scalar diverges at the throat, yielding a naked singularity described by Eq. (\ref{NS}) with $\omega=2$. On the other hand, when $\epsilon\neq 0$, it remains finite across all considered $n$ values. Additionally, we note that for $n>1$, the scalar $K$ has one minimum and maximum locally near the throat. The value of the Kretschmann scalar is a global maximum at the wormhole throat, increasing with the value of the power, $n$, while there is also a local maximum that increases with $n$, indicating that the stability of the wormhole solution decreases for large $n$. The local minimum shifts slightly with $n$ towards the throat but remains on the $n=1$ curve. Next, we will analyze the properties of the matter content associated with our wormhole solution (\ref{eq:lineCasimirAbelian}). 

\subsection{Source properties and stability analysis}

Now we are interested in studying the properties of the matter content responsible for forming our family of wormhole solutions (\ref{eq:lineCasimirAbelian}). 

We have observed that our distortion method does not affect the Casimir energy density $\rho(r)$, which remains with the same expression as in Eq. (\ref{eq:density}). However, the original Casimir radial pressure (\ref{eq:EoS}) changes to the following form
\begin{equation}
p_{r}^{(n)}(r)=\omega_{r}^{(n)}(r)\rho(r)=\omega\left[1-\left(\frac{r_{0}}{r}\right)^{n}\right]\rho(r).\label{prn}
\end{equation}

It is worth highlighting the connection between the original EoS and the modified equation that we have used here. Taking the mean value of $\omega_{r}^{(n)}$, we obtain 
\begin{equation}
\left\langle \omega_{r}^{(n)}\right\rangle =\lim_{r'\rightarrow\infty}\frac{\omega}{r'-r_{0}}\int_{r_{0}}^{r'}\left[1-\left(\frac{r_{0}}{r}\right)^{n}\right]dr=\omega,
\end{equation}which means that our approach recovers the standard relation between the density energy and the radial pressure from the Casimir effect, as expressed by their mean values. Specifically, when $\omega=2$, we obtain the Casimir relation (\ref{eq:EoS}):
\begin{equation}
\left\langle p_{r}^{(n)}\right\rangle =\left\langle \omega_{r}^{(n)}\right\rangle \rho=p_{r}^{\text{Cas}}.
\end{equation}

Another essential piece of information about the matter content comes from the lateral pressure $p_{\theta}$. Using the last Einstein equation (\ref{eq:gthetatheta}) and substituting our solutions (\ref{eq:bsolution}) and (\ref{eq:Phi_n}), we can write (with $\epsilon=\omega r_{0}^{n}$ and $r_{0}=2\lambda\kappa$):
\begin{equation}
p_{\theta}^{(n)}(r)=\omega_{\theta}^{(n)}(r)\rho(r),\label{pthetan}
\end{equation}where $\omega_{\theta}^{(n)}(r)$ has the following explicit form for some $n$-values:
\begin{eqnarray}
\omega_{\theta}^{(1)}(r)&=&\frac{\omega^{2}r_{0}(r_{0}-r)+\omega r(5r_{0}-4r)}{2r^{2}},\\
\omega_{\theta}^{(2)}(r)&=&\frac{\omega^{2}r_{0}(r_{0}-r)(r_{0}+r)^{2}+\omega r^{2}(7r_{0}^{2}-r_{0}r-4r^{2})}{2r^{4}},\\
\omega_{\theta}^{(3)}(r)&=&\frac{\omega^{2}r_{0}(r_{0}-r)(r_{0}^{2}+r_{0}r+r^{2})^{2}+\omega r^{3}(9r_{0}^{3}-r_{0}^{2}r-r_{0}r^{2}-4r^{3})}{2r^{6}}.
\end{eqnarray}
Note that for any finite value of $n$, the associated $\omega_{\theta}^{(n)}(r)$ is finite on the throat and vanishes when $r\rightarrow\infty$. Moreover, we can explicitly verify that the EoM (\ref{eq:g00}-\ref{eq:gthetatheta}) and the conservation law (\ref{eq:conservation}) are satisfied by the above expressions.

\begin{figure}[!h]
\begin{center}
\begin{tabular}{ccc}
\includegraphics[height=5cm]{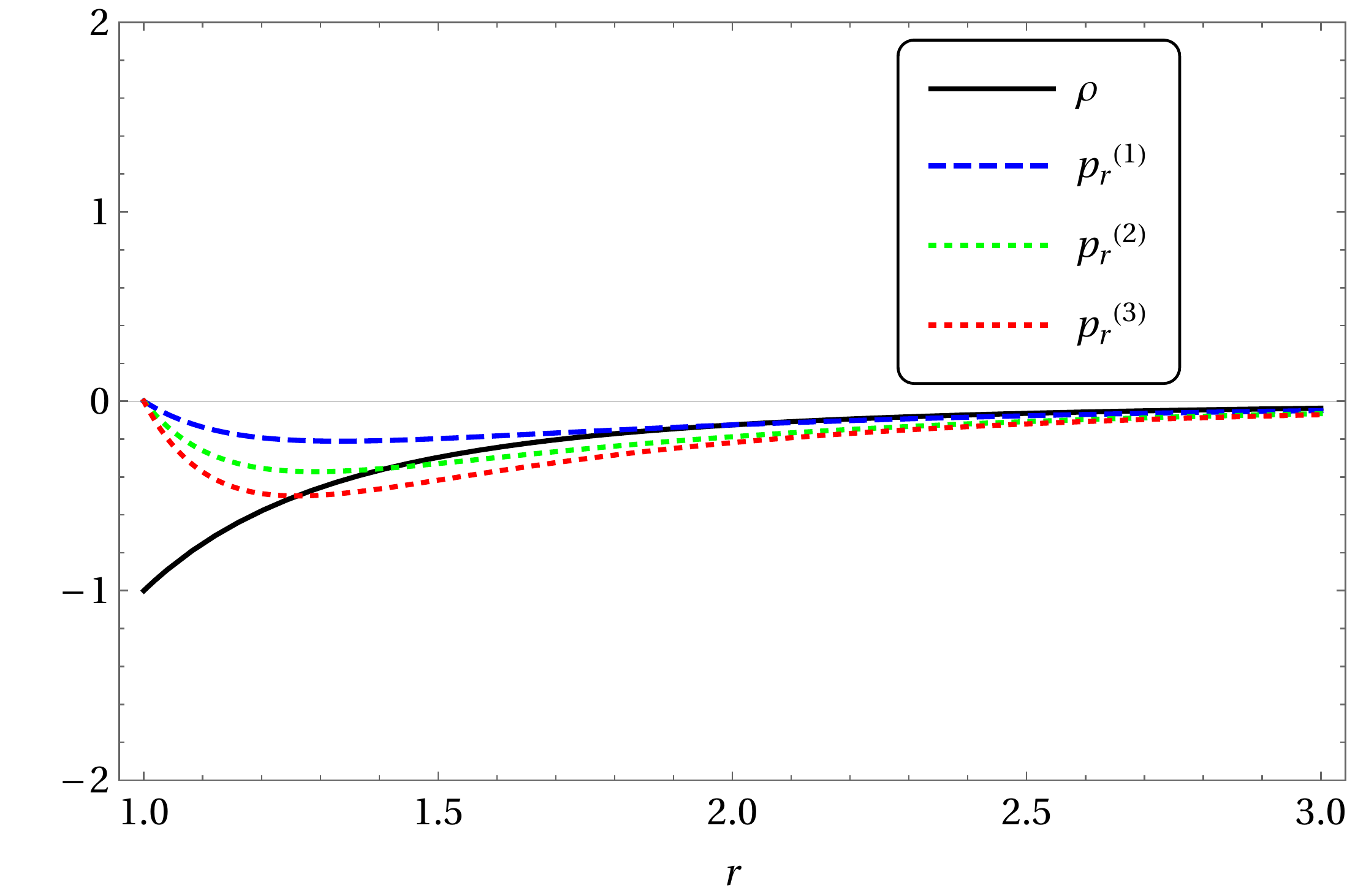}
\includegraphics[height=5cm]{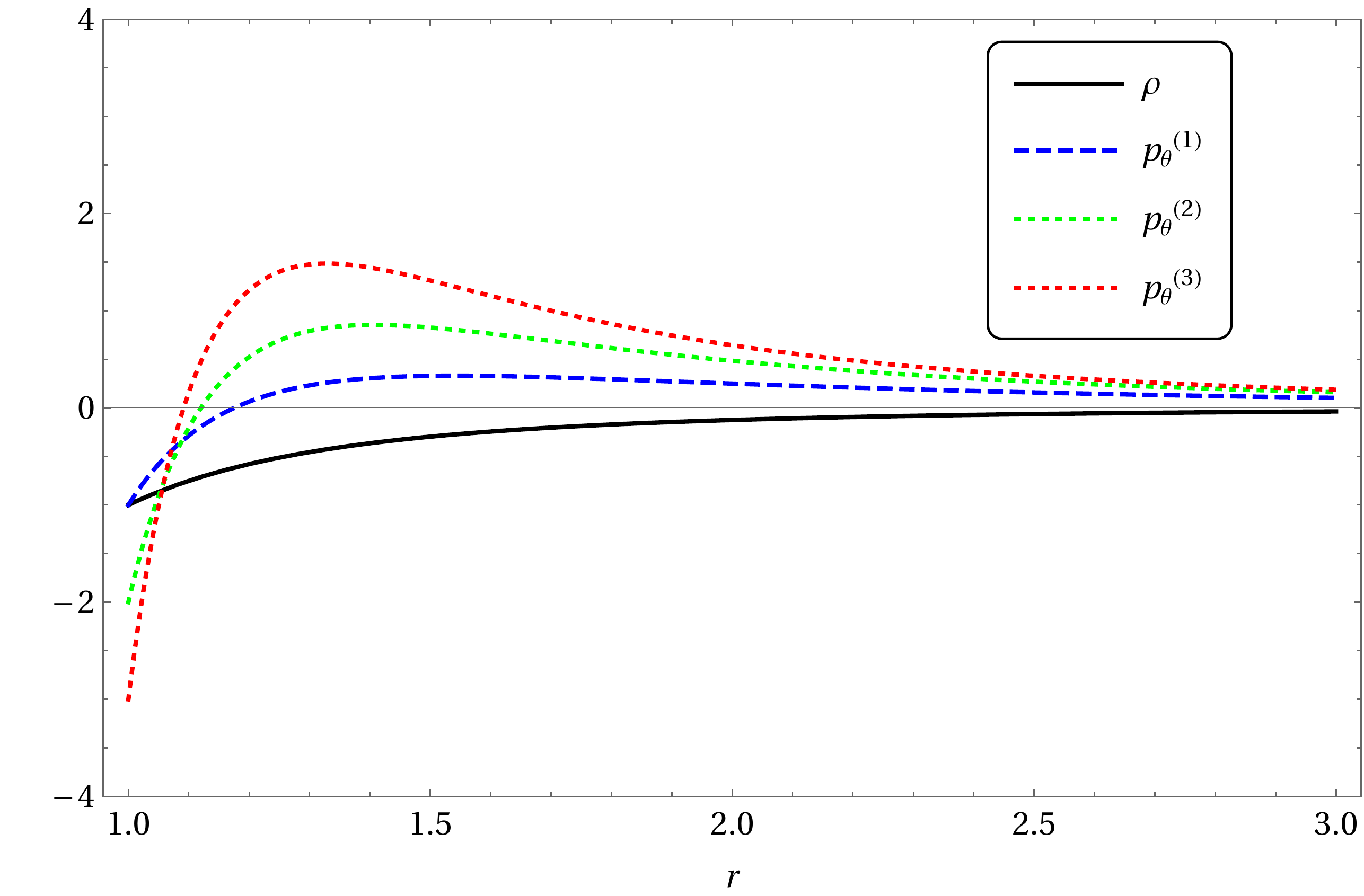}\\ 
(a) \hspace{8 cm}(b)\\
\includegraphics[height=5cm]{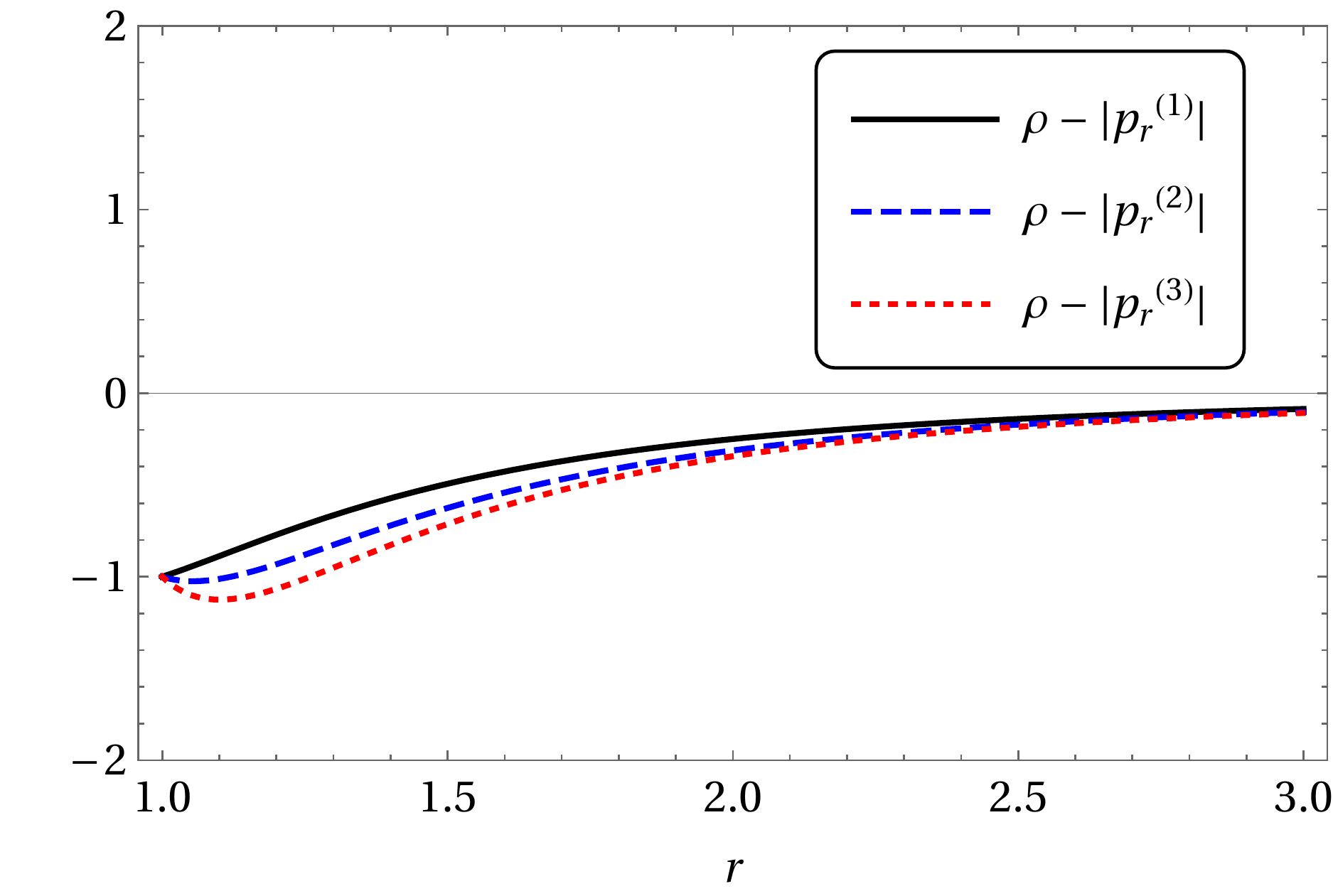}
\includegraphics[height=5cm]{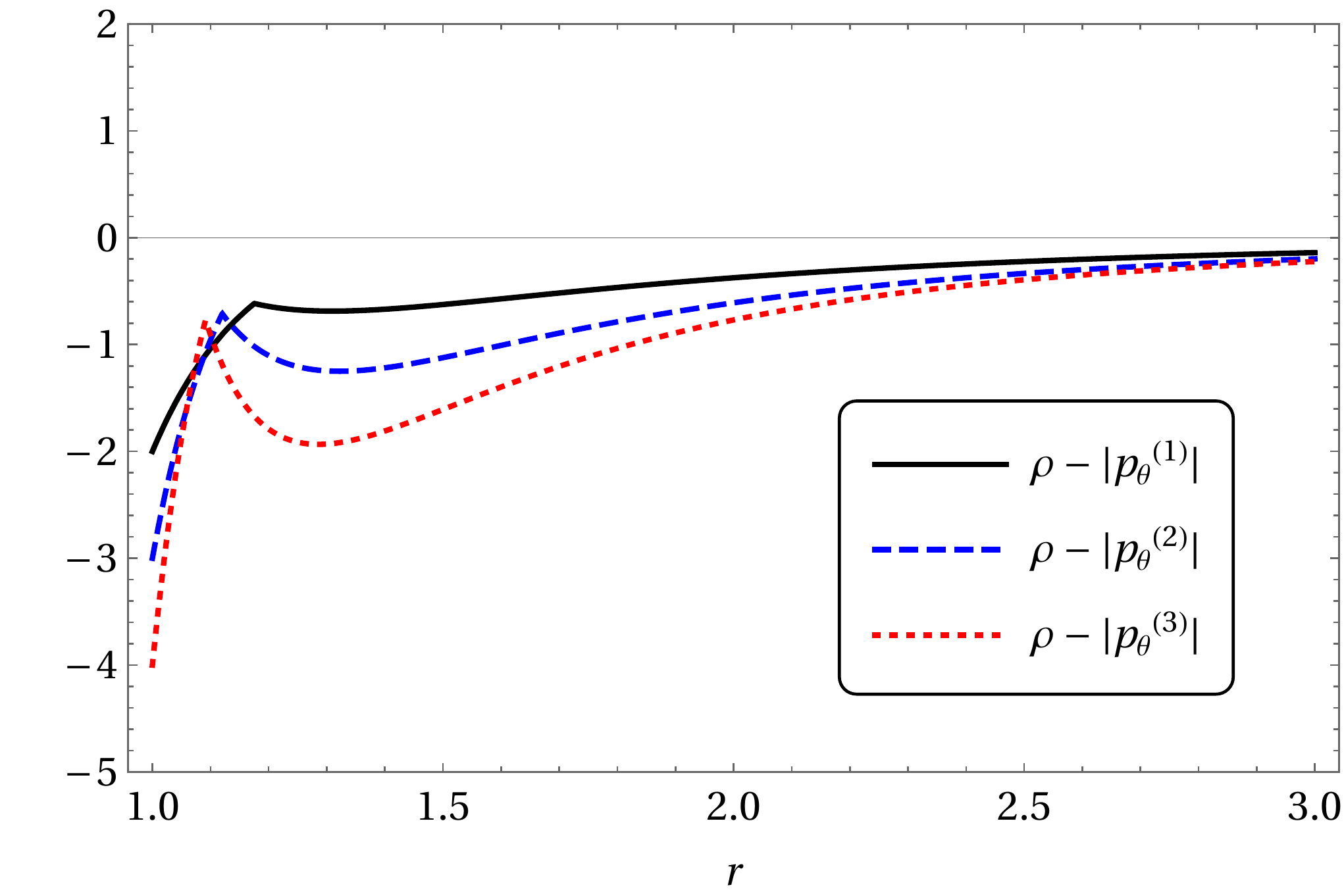}\\ 
(c) \hspace{8 cm}(d)
\end{tabular}
\end{center}
\caption{
Density and radial pressures (a), density and lateral pressures (b), and combination of density and pressures (c) and (d) in order to verify energy conditions, namely, DEC, considering $n=1,2,3$, $\omega=2$, and $r_{0}=\lambda=1$, in Planck units.
\label{fig2}}
\end{figure}

Figure \ref{fig2} provides a comprehensive visualization of the energy density, radial and lateral pressures, and their various combinations to aid in the analysis of energy conditions. Specifically, the null energy condition (NEC), which states that the energy density plus any of the pressures cannot be negative ($\rho+p_i\geq 0$), and the dominant energy condition (DEC), which states that the energy density must be greater than or equal to the absolute value of any of the pressures ($\rho-|p_i|\geq 0$), are examined. It is worth noting that these conditions are not satisfied throughout the domain, as evidenced by the violation of the NEC and DEC. Furthermore, the strong energy condition (SEC), which requires the NEC to hold true and $\rho+\sum_i p_i\geq 0$, is also not satisfied since NEC is violated.

\subsubsection{Stability from sound velocity}

\begin{figure}[!]
\begin{center}
\includegraphics[height=6.5cm]{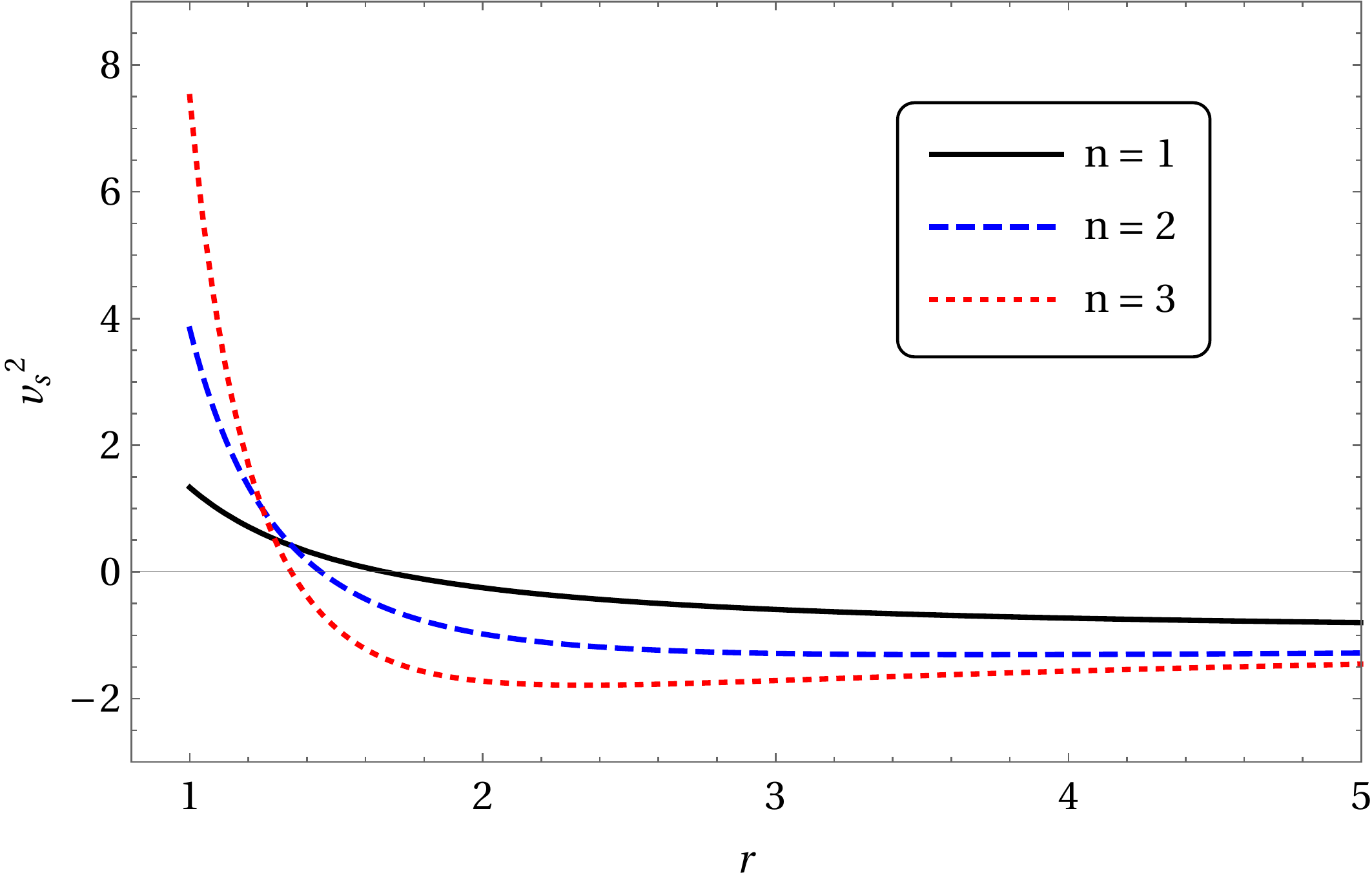}
\caption{Squared sound velocity as a function of the radial coordinate, for the noninteracting case, considering $\omega=2$ and $r_0=1$.}\label{Sound2}
\end{center}
\end{figure}

In order to evaluate the stability of the Casimir wormhole family given by Eq. (\ref{eq:lineCasimirAbelian}), we must firstly analyze the condition described by the expression \cite{Capozziello:2020zbx, Capozziello:2022zoz}:
\begin{equation}
v_s^2(r)=\frac{1}{2}\left[\frac{d(p_r+p_{\theta})}{d\rho}\right]=\frac{1}{2}\left[\frac{p_r'(r)+p_{\theta}'(r)}{\rho'(r)}\right]\geq 0,\label{Sound}
\end{equation}
with $v_s$ representing the sound velocity in the medium, where we take the average pressure between the lateral and radial ones in 2+1 dimensions, which are given by Eqs. (\ref{prn}) and (\ref{pthetan}). As previously demonstrated, the effective state parameter associated with the wormhole source is expressed by the function $\omega^{(n)}(r)=2[1-(r_0/r)^n]$, which must be shifted from value $\omega=2$ to enable the traversability of the wormhole. Thus, for $n=1$, the sound velocity takes the form
\begin{equation}
 v_{s}^{2(1)}(r)=\omega^{2}\left(\frac{5r_{0}^{2}}{12r^{2}}-\frac{r_{0}}{3r}\right)+\omega\left(\frac{r_{0}}{r}-\frac{1}{2}\right), 
\end{equation}such that it is independent of $\lambda$, and its value at the throat does not change with different choices for $r_{0}$. The same behavior is observed for all $n$ integer finite. The stability of the solutions is predominantly assured near the wormhole throat, as depicted in Figure \ref{Sound2}. 
Although stability is guaranteed, it is essential to acknowledge that the system is considered unphysical due to the sound speed consistently surpassing the speed of light near the throat, regardless of the parameter values. However, it is noteworthy that a superluminal velocity may not necessarily be forbidden in this context, given the presence of exotic matter responsible for the wormhole's formation \cite{Ellis:2007ic}. Nevertheless, we can address this concern in the interacting case by carefully selecting appropriate parameter values.

\subsubsection{Stability from TOV equation}

The stability of compact objects, such as stellar structures, can be also examined through the Tolman-Oppenheimer-Volkov (TOV) equation, initially introduced in the context of neutron stars \cite{Tolman:1939jz,Oppenheimer:1939ne}. Hence, the stability conditions for wormholes could be investigated by employing an equilibrium condition obtained from the TOV equation \cite{Gorini:2008zj,Kuhfittig:2020fue,Mustafa:2021bfs,Sokoliuk:2021rtv}. Combining the Morris-Thorne metric ansatz (\ref{eq:metricAnsatz}) with the conservation law of  stress-energy tensor $\nabla_{\mu}T^{\mu\nu}=0$, we obtain the following TOV equation in a $2+1$ dimensional spacetime:
\begin{equation}
\frac{dp_{r}}{dr}+\Phi'(\rho+p_{r})+\frac{1}{r}(p_{r}-p_{\theta})=0.\label{TOV1}
\end{equation}
The equilibrium state of the wormhole structure is determined by three terms in this equation, namely, the hydrostatic force ($\mathcal{F}_{h}$), the gravitational force ($\mathcal{F}_{g}$), and the anisotropic force ($\mathcal{F}_{a}$), defined as follows \cite{Kuhfittig:2020fue}:
\begin{eqnarray}
\mathcal{F}_{h}=-\frac{dp_{r}}{dr},\ \ \ \ \ \mathcal{F}_{g}=-\Phi'(\rho+p_{r}),\ \ \ \ \ \mathcal{F}_{a}=\frac{1}{r}(p_{\theta}-p_{r}).
\end{eqnarray}Thus, Eq. (\ref{TOV1}) takes the simple form given by:
\begin{equation}
\mathcal{F}_{h}+\mathcal{F}_{g}+\mathcal{F}_{a}=0.\label{TOV2}
\end{equation}
It is straightforward to verify that our class of traversable wormhole solutions satisfies the TOV equation for all finite values of $n$. For instance, when $n=1$, and employing Eqs (\ref{eq:density}), (\ref{eq:solTWn=00003D1}), (\ref{prn}), and (\ref{pthetan}), we obtain the following expressions for the forces:
\begin{eqnarray}
\mathcal{F}_{h}	&=&-\frac{\lambda\omega(3r-4r_{0})}{r^{5}},\\
\mathcal{F}_{g}	&=&-\frac{\lambda\omega(r_{0}r+\omega r_{0}r-\omega r_{0}^{2})}{2r^{6}},\\
\mathcal{F}_{a}	&=&\frac{\lambda\omega(6r^{2}-7r_{0}r++\omega r_{0}r-\omega r_{0}^{2})}{2r^{6}},
\end{eqnarray}such that the equilibrium condition (\ref{TOV2}) is satisfied for all $r$, and confirming it as a stable wormhole.

\begin{figure}[!h]
\begin{center}
\begin{tabular}{ccc}
\includegraphics[height=5.1cm]{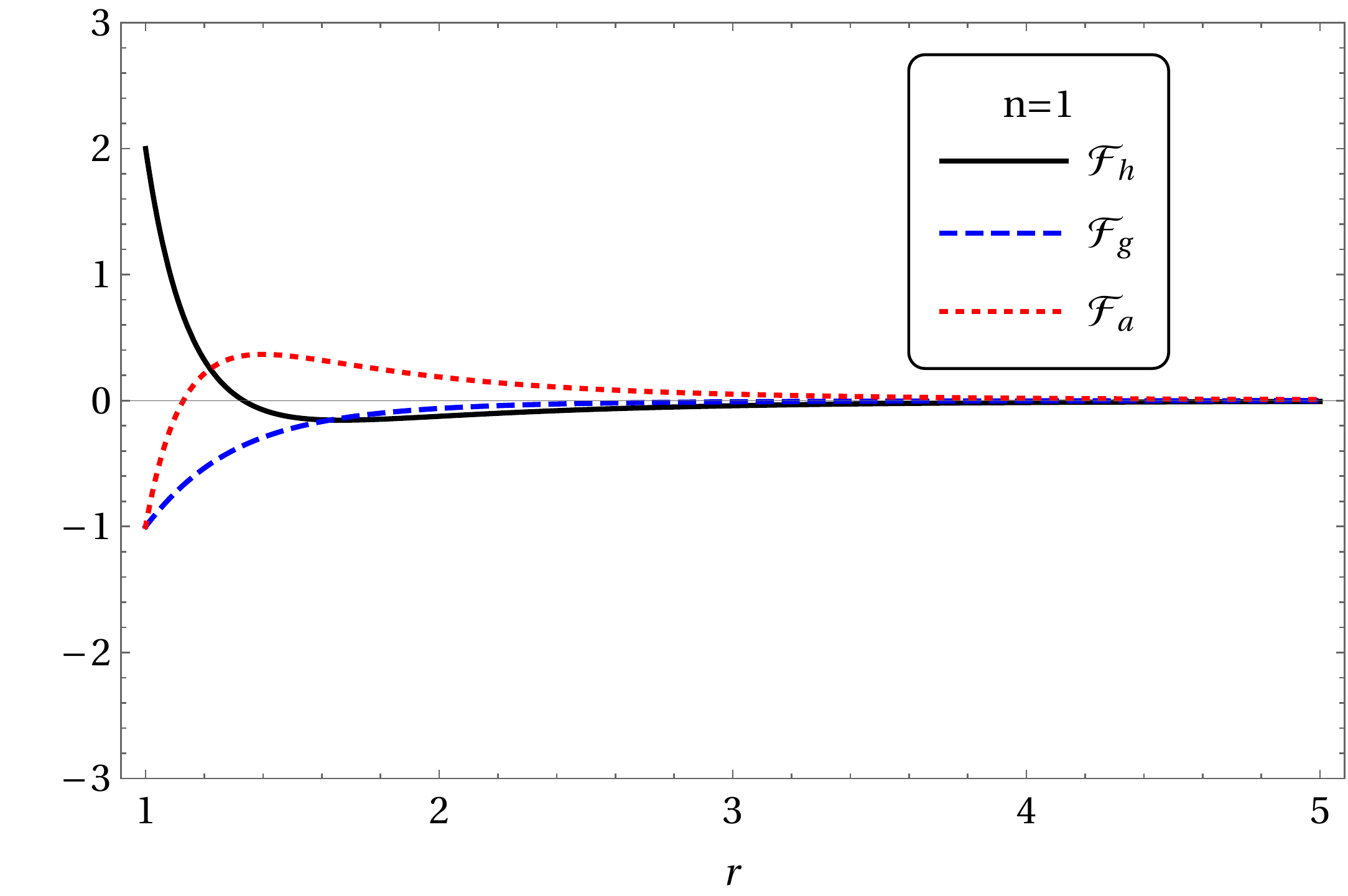}
\includegraphics[height=5cm]{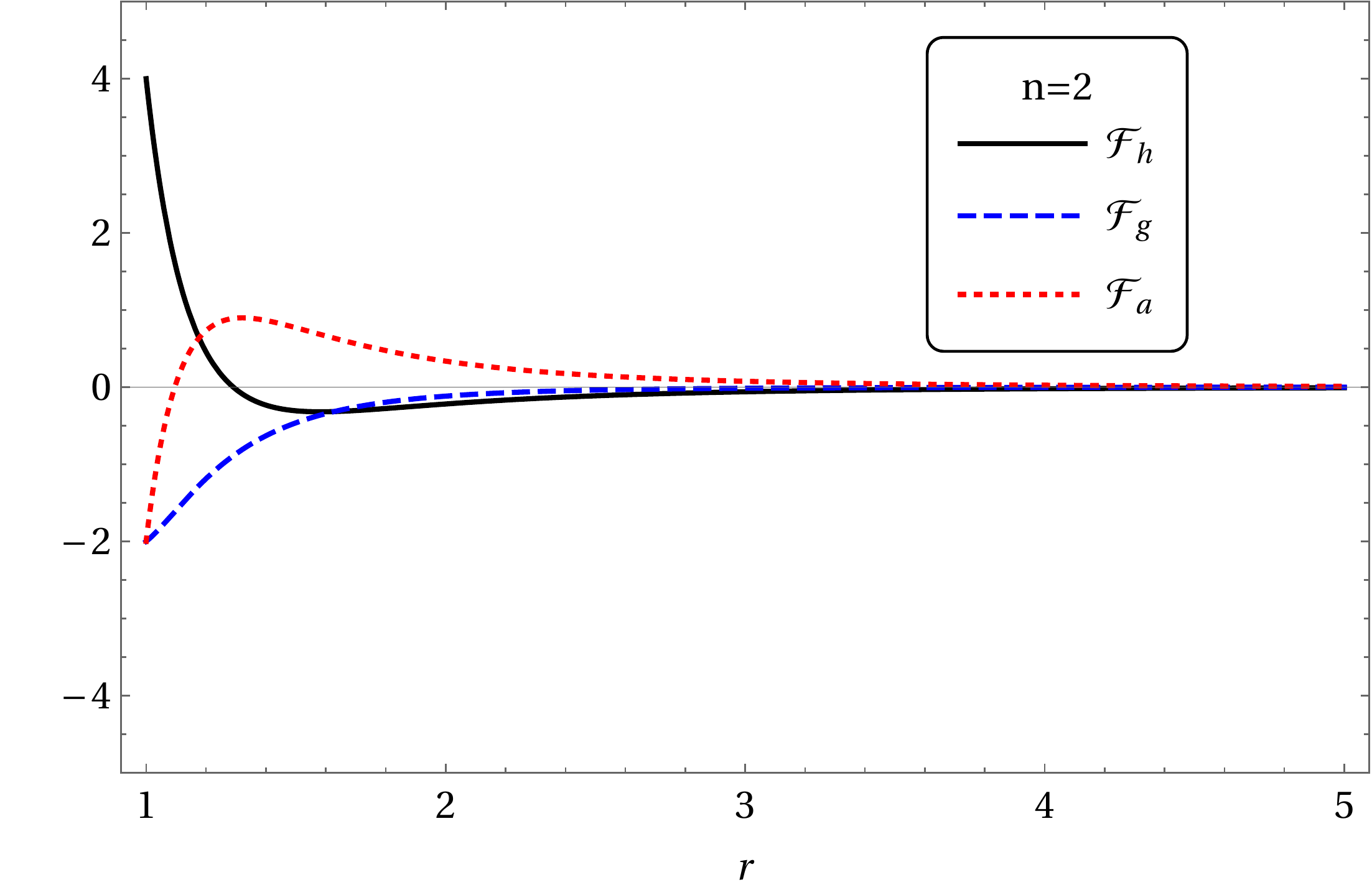}\\ 
(a) \hspace{8 cm}(b)
\end{tabular}
\end{center}
\caption{
The graphical representation of $\mathcal{F}_{h}$, $\mathcal{F}_{g}$, and $\mathcal{F}_{a}$ as functions of the radial coordinate $r$, with $\omega=2$ and $\lambda=r_{0}=1$, is shown in (a) for $n=1$ and in (b) for $n=2$.\label{figForce}}
\end{figure}

The profiles of $\mathcal{F}_{h}$, $\mathcal{F}_{g}$, and $\mathcal{F}_{a}$ for our wormhole solutions with $n=1$ and $n=2$ are depicted in Fig. \ref{figForce}(a) and \ref{figForce}(b), respectively, with the assigned values of the free parameters as $\omega=2$ and $\lambda=r_0=1$. We can observe that the hydrostatic force dominates over the gravitational and anisotropic forces. For both cases, $\mathcal{F}_{h}$ takes positive values near the throat, while $\mathcal{F}_{g}$ and $\mathcal{F}_{a}$ are negative, clearly indicating that to maintain the system in an equilibrium state, the hydrostatic force is balanced by the combined effect of gravitational and the anisotropic forces. Furthermore, the magnitude of the forces near the throat increases as $n$ grows. This can be interpreted as an effect of increasing curvature with the value of $n$, as verified earlier in our analysis of the Kretschmann scalar.

\section{Interacting (confined) case\label{sec3}}

The Casimir energy of two parallel wires in the $(2+1)$-dimensional
Yang-Mills theory was analyzed using both a Hamiltonian approach \cite{Karabali:2018ael}
and lattice simulation \cite{Chernodub:2018pmt}. Despite the distinct analytical expressions for the renormalized zero-point energy found in the cited works, they share common features: Both results show graphs very close to the Casimir energy profile, and a massive quantity exponentially suppresses the Casimir energy. This emergent mass scale may be related to the glueball masses and influences the confinement-deconfinement transition of the gluonic fields between the wires \cite{Pasechnik:2016wkt}.

Thus, in the context of our present work, the Casimir energy's infrared damping may affect the wormhole solutions behavior, resulting in new physical effects. These implications have not yet been evaluated in the literature and will be the main focus of this section.

Effectively, the lattice calculation of the Casimir energy for the
$SU(N)$ gauge theory was fitted by the following expression \cite{Chernodub:2018pmt}
\begin{equation}
\frac{\mathcal{E}}{L}=-\mbox{dim}G\frac{\zeta(3)}{16\pi}\frac{\left(R\sqrt{\sigma}\right)^{-\nu}}{R^{2}}e^{-M_{\text{Cas}}R},\label{eq:CasimirYM1}
\end{equation}
where $L$ is the length of the wire, $R$ is the distance between
them, and $\sigma$ is the tension of the confining (fundamental)
Yang-Mills string at zero temperature. The exponent $\nu$ represents
an anomalous dimension of the Casimir potential at short distances,
and the quantity $M_{\text{Cas}}$ represents the effective Casimir
mass associated with the nonperturbative mass gap at large distances. The free parameters $\nu$ and $M_{\text{Cas}}$ can be tuned to obtain
the best lattice fit. In the case of $SU(2)$, where $\mbox{dim}G=3$,
the best fit for lattice simulations was obtained for $\nu=0.05(2)$
and $M_{\text{Cas}}=1.38(3)\sqrt{\sigma}$ . As noted by the authors
in \cite{Chernodub:2018pmt}, the value of $M_{\text{Cas}}$ is considerably lower
than the minimum value estimated for the glueball mass, which is around
$4.7\sqrt{\sigma}$ \cite{Karabali:2018ael, Teper:1998te,Athenodorou:2016ebg}.

Adopting the procedure of the previous sections, let us promote the
distance between the wires to the radial coordinate $r$ so that the
Casimir energy density $\rho(r)$ and the associated radial pressure
$p_{r}(r)$ can be calculated as follows: 

\begin{equation}
\rho(r)=\frac{\mathcal{E}}{Lr}=-\lambda\frac{\left(r\sqrt{\sigma}\right)^{-\nu}}{r^{3}}e^{-M_{\text{Cas}}r},\label{eq:densityYM}
\end{equation}
with $\lambda=\mbox{dim}G \zeta(3)/16\pi$, and 
\begin{align}
p_{r}(r) & =\frac{F}{L}=-\frac{1}{L}\frac{d\mathcal{E}}{dr},\nonumber \\
 & =-\lambda\frac{(2+\nu+M_{\text{Cas}}r)\left(r\sqrt{\sigma}\right)^{-\nu}}{r^{3}}e^{-M_{\text{Cas}}r},\label{eq:pressureYM}
\end{align}
where we write $p_{r}(r)=\omega(r)\rho(r)$ with $\omega(r)=2+\nu+M_{\text{Cas}}r.$ It is worth noting that
when the interactions are turned off, $M_{\text{Cas}}=0$ and $\nu=0$,
and the expressions (\ref{eq:densityYM})-(\ref{eq:pressureYM}) reduce
to their tree-level form (\ref{eq:density})-(\ref{eq:EoS}), respectively.

Substituting (\ref{eq:densityYM}) into (\ref{eq:g00}), we obtain the
shape function of the Casimir-Yang-Mills wormhole as 
\begin{equation}
b(r)=r+2\kappa\lambda\sigma^{-\frac{\nu}{2}}M_{\text{Cas}}^{\nu+1}r\left[\Gamma(-\nu-1,M_{\text{Cas}}r)-\Gamma(-\nu-1,M_{\text{Cas}}r_{0})\right],\label{eq:brYM}
\end{equation}
where $\Gamma(a,z)$ is the incomplete gamma function $\Gamma(a,z)=\int_{z}^{\infty}dtt^{a-1}e^{-t}$.
Naturally, the constant of integration was fixed such that the throat
condition $b(r_{0})=r_{0}$ is satisfied.

\begin{figure}[!h]
\begin{center}
\includegraphics[height=7.3cm]{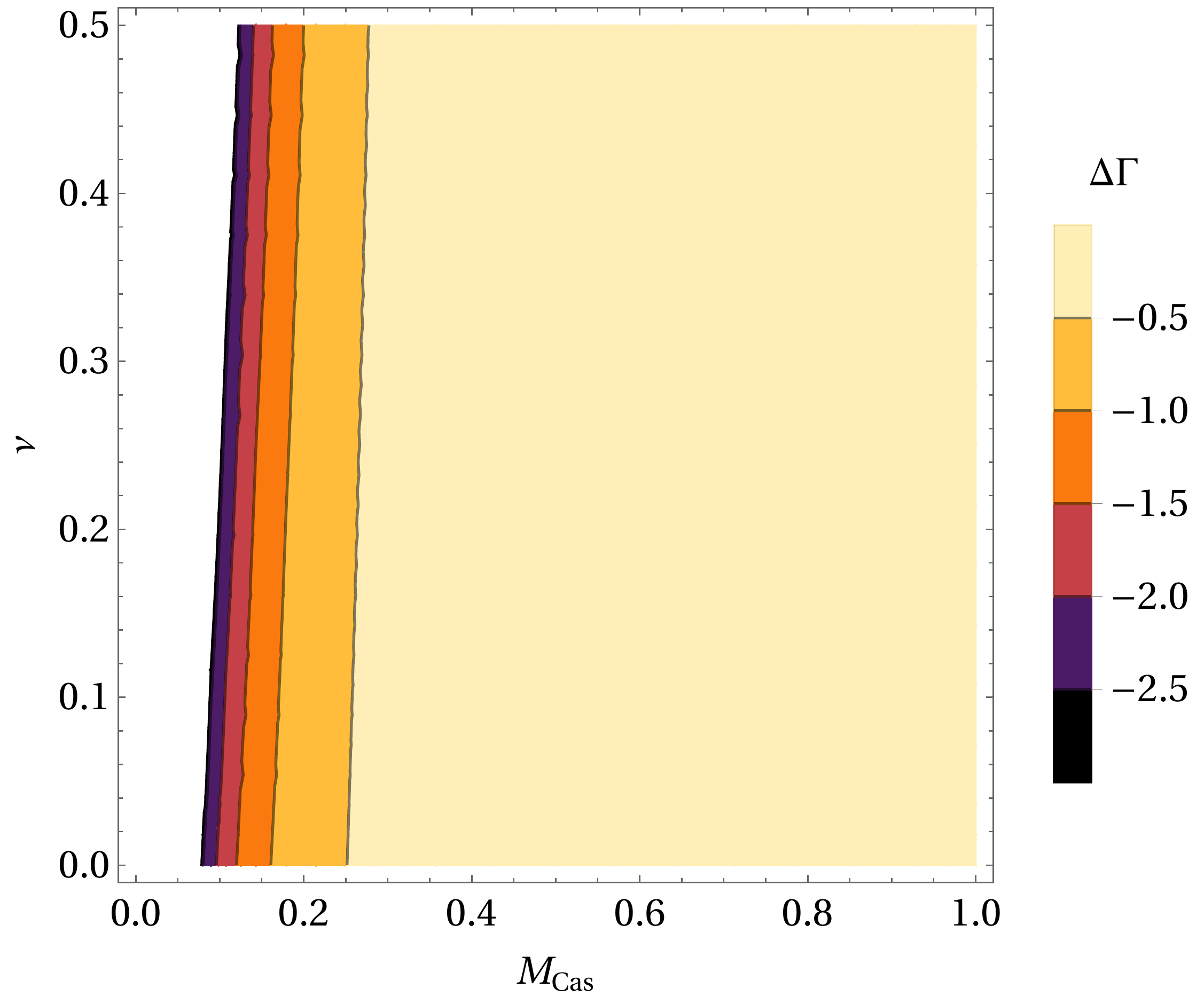}
\caption{Influence of the parameters $\nu$ and $M_{\text{Cas}}$ on the inequality (\ref{eq:desi}). In this graphic, we adopted $r=4$ and $r_0=2$ (in natural units $\hbar=c=\tilde{G}=1$).\label{figDeltaGama1}}
\end{center}
\end{figure}

Our goal now is to investigate how the Casimir shape function is affected
by non-abelian corrections, which are characterized by the parameters
$\nu$ and $M_{\text{Cas}}$. First, we may note that in the $\nu\rightarrow0$
and $M_{\text{Cas}}\rightarrow0$ limits, the expression (\ref{eq:brYM})
reduces to
\begin{equation}
\lim_{\begin{smallmatrix}\nu\to0\\
M_{\text{Cas}}\rightarrow0
\end{smallmatrix}}b(r)=2\kappa\lambda+\left(1-\frac{2\kappa\lambda}{r_{0}}\right)r,
\end{equation}
agreeing with our original result (\ref{eq:bsolution}). Moreover,
the flare-out condition $(i)$ is satisfied for all range of parameters,
namely,
\begin{equation}
\frac{b-rb'}{b^{2}}=\frac{2\kappa\lambda\sigma^{\nu/2}r^{-\nu-2}e^{-M_{\text{Cas}}r}}{\left[2\kappa\lambda M_{\text{Cas}}^{\nu+1}\left(\Gamma(-\nu-1,M_{\text{Cas}}r)-\Gamma(-\nu-1,M_{\text{Cas}}r_{0})\right)+\sigma^{\nu/2}\right]^{2}}>0,
\end{equation}
and on the throat
\begin{equation}
b'(r_{0})=1-2\kappa\lambda\sigma^{-\frac{\nu}{2}}r_{0}^{-\nu-1}e^{-M_{\text{Cas}}r_{0}}<1,
\end{equation}
since $\kappa$, $\lambda$, $\sigma$, $\nu$, and $M_{\text{Cas}}$ are
positive real quantities. Also, the property $(ii)$ $1-b/r\geq0$
holds true if the inequality
\begin{equation}
\Delta\Gamma\equiv\Gamma(-\nu-1,M_{\text{Cas}}r)-\Gamma(-\nu-1,M_{\text{Cas}}r_{0})\leq0,\label{eq:desi}
\end{equation}
is satisfied, which is always the case for $r\geq r_{0}$. The influence
of the $\nu$ and $M_{\text{Cas}}$ parameters on the inequality (\ref{eq:desi})
is indicated in Fig. \ref{figDeltaGama1}, where we take $r=4$ and $r_{0}=2$. As we
can see, for large values of $M_{\text{Cas}}$, the difference in
(\ref{eq:desi}) goes to zero.

The asymptotic flatness condition, $\lim_{r\rightarrow\infty}(1-b(r)/r)=1$,
implies that 
\begin{equation}
1-2\kappa\lambda\sigma^{-\frac{\nu}{2}}M_{\text{Cas}}^{\nu+1}\Gamma(-\nu-1,M_{\text{Cas}}r_{0})=0,\label{eq:flatness}
\end{equation}
which involves solving a transcendental equation for $r_{0}$. This
equation does not have an analytic solution for arbitrary values of
$\nu$, $\sigma$, and $M_{\text{Cas}}$. To get some insight into
the effect of Yang-Mills corrections on the throat radius $r_{0}$,
we can expand the equation (\ref{eq:flatness}) for small values of
$\nu$ and $M_{\text{Cas}}$. The result can be written as 
\begin{equation}
r_{0}-2\kappa\lambda\left\{ 1-M_{\text{Cas}}r_{0}\left[1-\gamma_{\text{E}}-\ln\left(M_{\text{Cas}}r_{0}\right)\right]-\frac{\nu}{2}\left[2+\ln\left(\sigma r_{0}^{2}\right)\right]\right\} =0,\label{eq:roapprox}
\end{equation}
where $\gamma_{\text{E}}$ is the Euler-Mascheroni constant, with
numerical value $\gamma_{\text{E}}\approx0.577216$. Notice that we
recover our tree-level result $r_{0}^{\text{\mbox{tree}}}=2\kappa\lambda$
for $\nu=0$ and $M_{\text{Cas}}=0$. Furthermore, the non-abelian interaction causes a decrease in the throat radius, which may eventually
result in its closure.

\begin{figure}[!]
\begin{center}
\includegraphics[height=8cm]{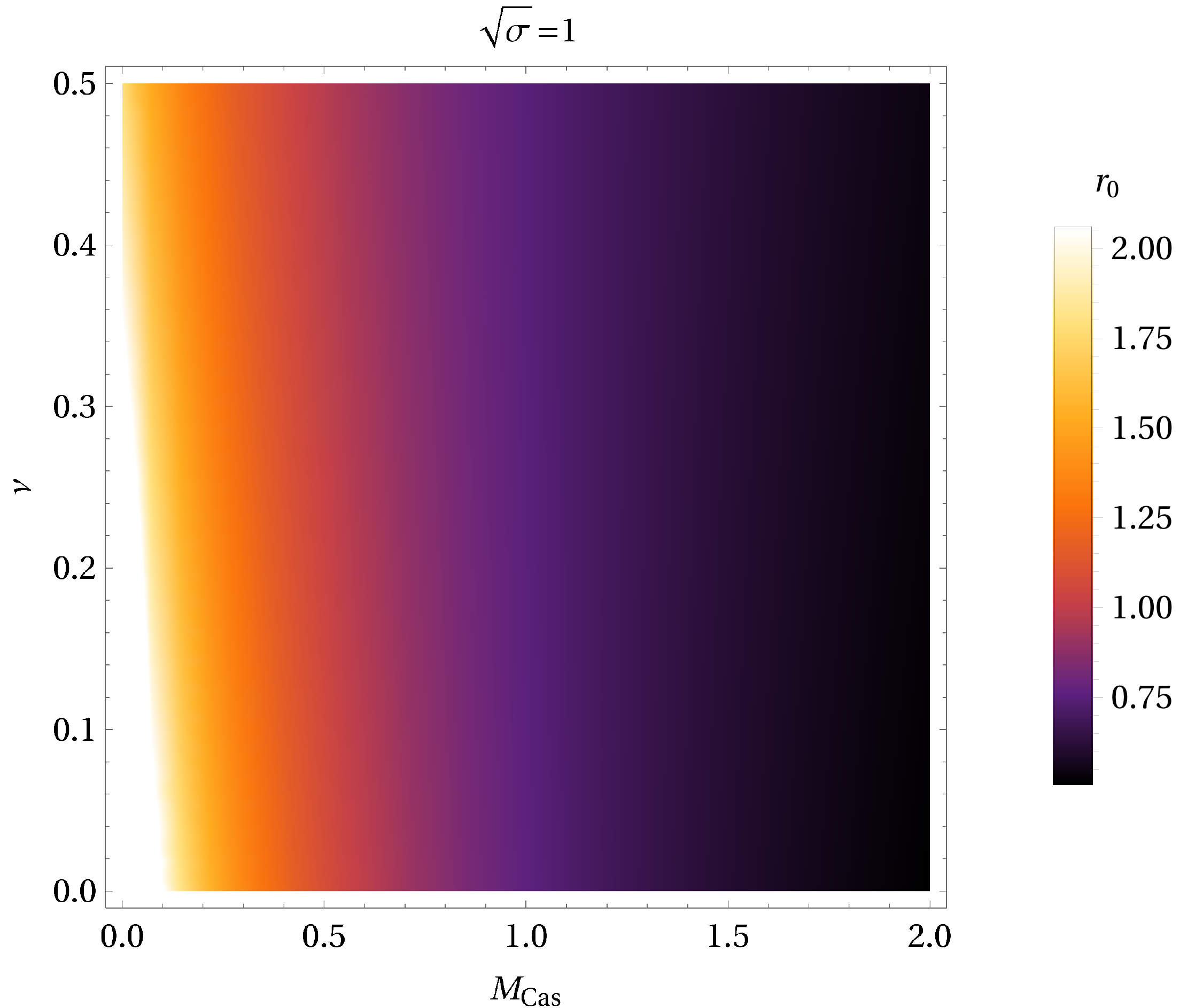}
\caption{Numerical solution for (\ref{eq:flatness}) with respect to the parameters $\nu$ and $M_{\text{Cas}}$. In this graphic, we adopted $\sqrt{\sigma}=1$,  $\lambda=3\zeta(3)/16\pi$ and $\kappa=8\pi$ (in natural units $\hbar=c=\tilde{G}=1$).}\label{running1}
\end{center}
\end{figure}

\begin{figure}[!]
\begin{center}
\includegraphics[height=7cm]{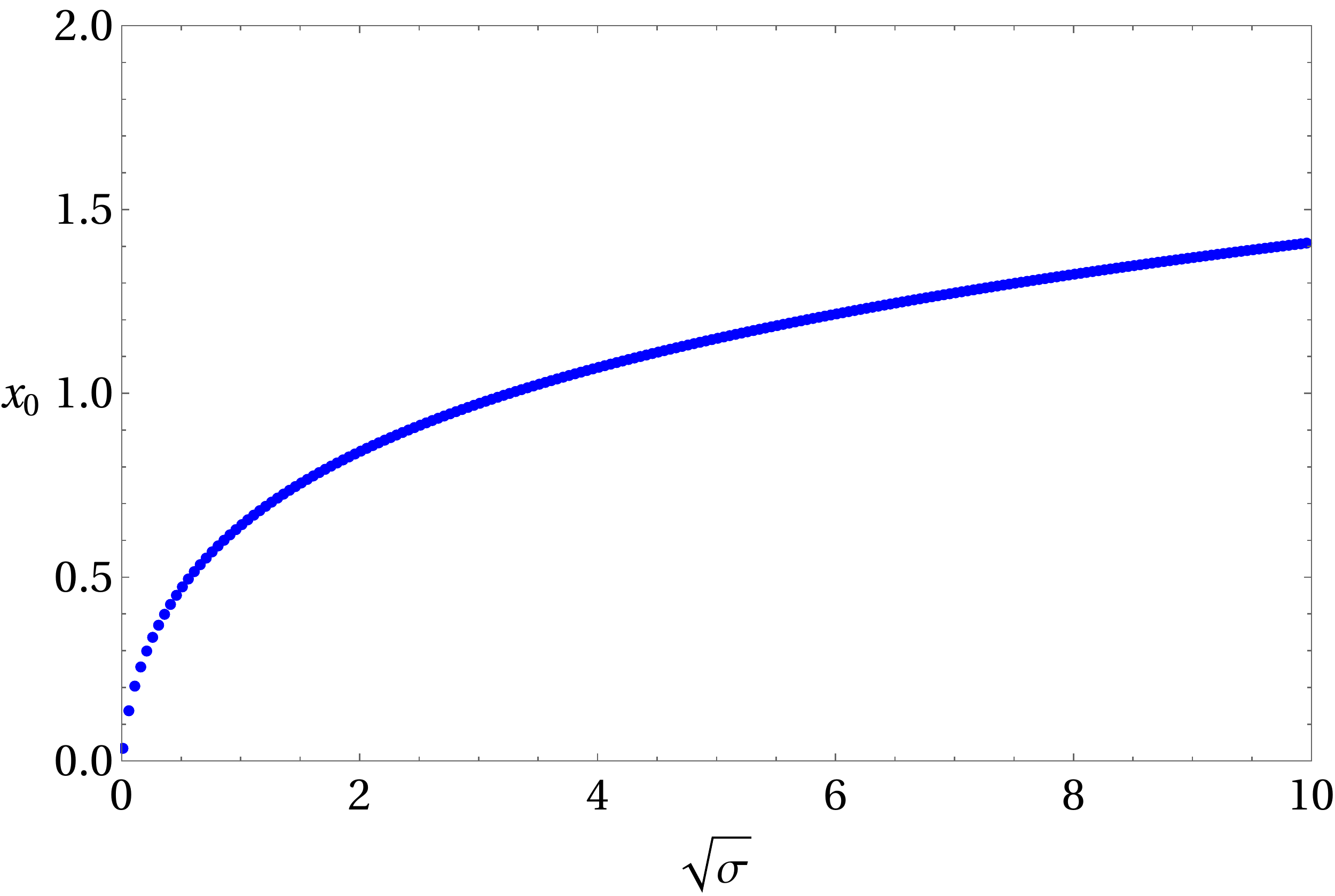}
\caption{Numerical solution for (\ref{eq:flatness}) with respect to the parameter $\sqrt{\sigma}$. In this plot, we use $\lambda=3\zeta(3)/16\pi$, $\kappa=8\pi$ and fix $\nu=0.05$ and $M_{\text{Cas}}=1.38\sqrt{\sigma}$ (in natural units $\hbar=c=\tilde{G}=1$).}\label{running2}
\end{center}
\end{figure}

\begin{figure}[!]
\begin{center}
\begin{tabular}{ccc}
\includegraphics[height=4.5cm]{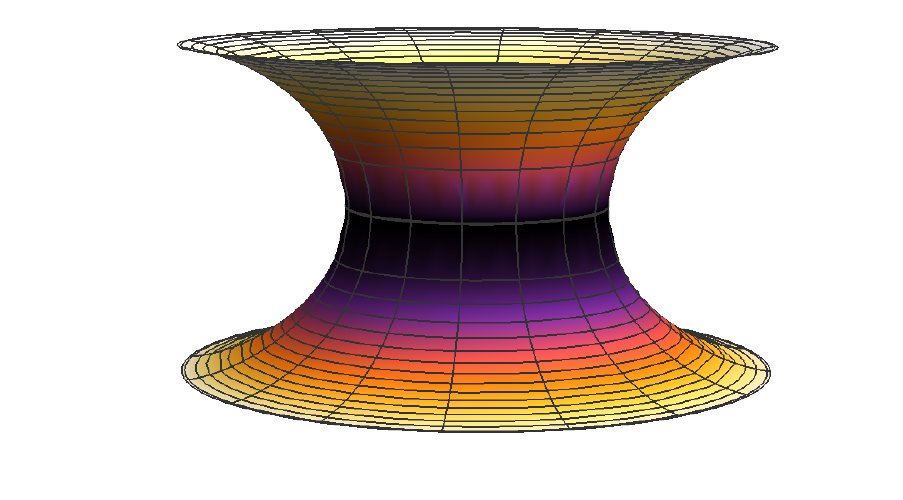}
\includegraphics[height=4.5cm]{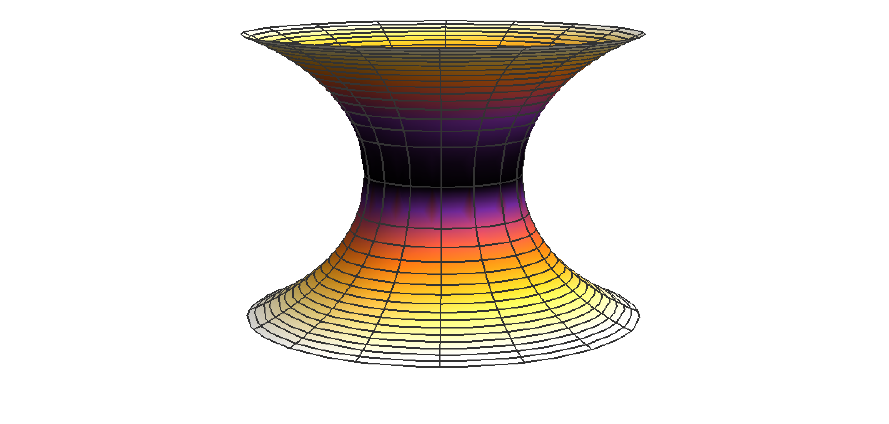}\\ 
(a) \hspace{8 cm}(b)\\
\end{tabular}
\end{center}
\caption{Embedded shapes of the Casimir-Yang-Mills wormhole: with $M_{\text{Cas}}=0.001$ and $r_{0}=3.1841$ (a), with $M_{\text{Cas}}=0.1$ and $r_{0}=1.9602$ (b). In both cases we fix $\nu=0.05$ and $\sqrt{\sigma}=1$. Also, we assumed $\lambda=3\zeta(3)/16\pi$ and $\kappa=8\pi$ (in natural units $\hbar=c=\tilde{G}=1$).\label{figembedding}}
\end{figure}

In order to determine the dependence of the throat radius $r_{0}$ on the Yang-Mills parameters, namely, the string tension $\sigma$, anomalous dimension $\nu$, and the Casimir mass $M_{\text{Cas}}$, we consider two separate cases. First, we fix $\sqrt{\sigma}=1$ and consider the running of $\nu$ and $M_{\text{Cas}}$. The resultant dependence of $r_{0}$ with respect to $\nu$ and $M_{\text{Cas}}$ is depicted in Fig. \ref{running1}, where we adopt $\lambda=3\zeta(3)/16\pi$ and $\kappa=8\pi$. As we can see, when the parameters increase, the radius $r_0$ decreases, as expected from our perturbative result (\ref{eq:roapprox}). Next, we assume the best-fit values of the parameters $\nu$ and $M_{\text{Cas}}$ for the $SU(2)$ gauge group as calculated via lattice techniques in Ref. \cite{Chernodub:2018pmt}. Specifically, we set $\nu=0.05$ and $M_{\text{Cas}}=1.38\sqrt{\sigma}$, and introduce the variable $x_{0}=r_{0}\sqrt{\sigma}$ to numerically solve Eq. (\ref{eq:flatness}). The numerical results are presented in Fig. \ref{running2}. The graph shows that as $\sqrt{\sigma}$ increases, the value of $x_{0}$ increases gradually.  This outcome can be attributed to the stretching of the wormhole throat, induced by the progressive increase in string tension, which is designed to release the gluons from their confined state. The dependence of the throat radius on the Casimir mass is also visualized by examining the embedding geometry, as illustrated in Figure \ref{figembedding}.

\begin{figure}[!]
\includegraphics[height=7cm]{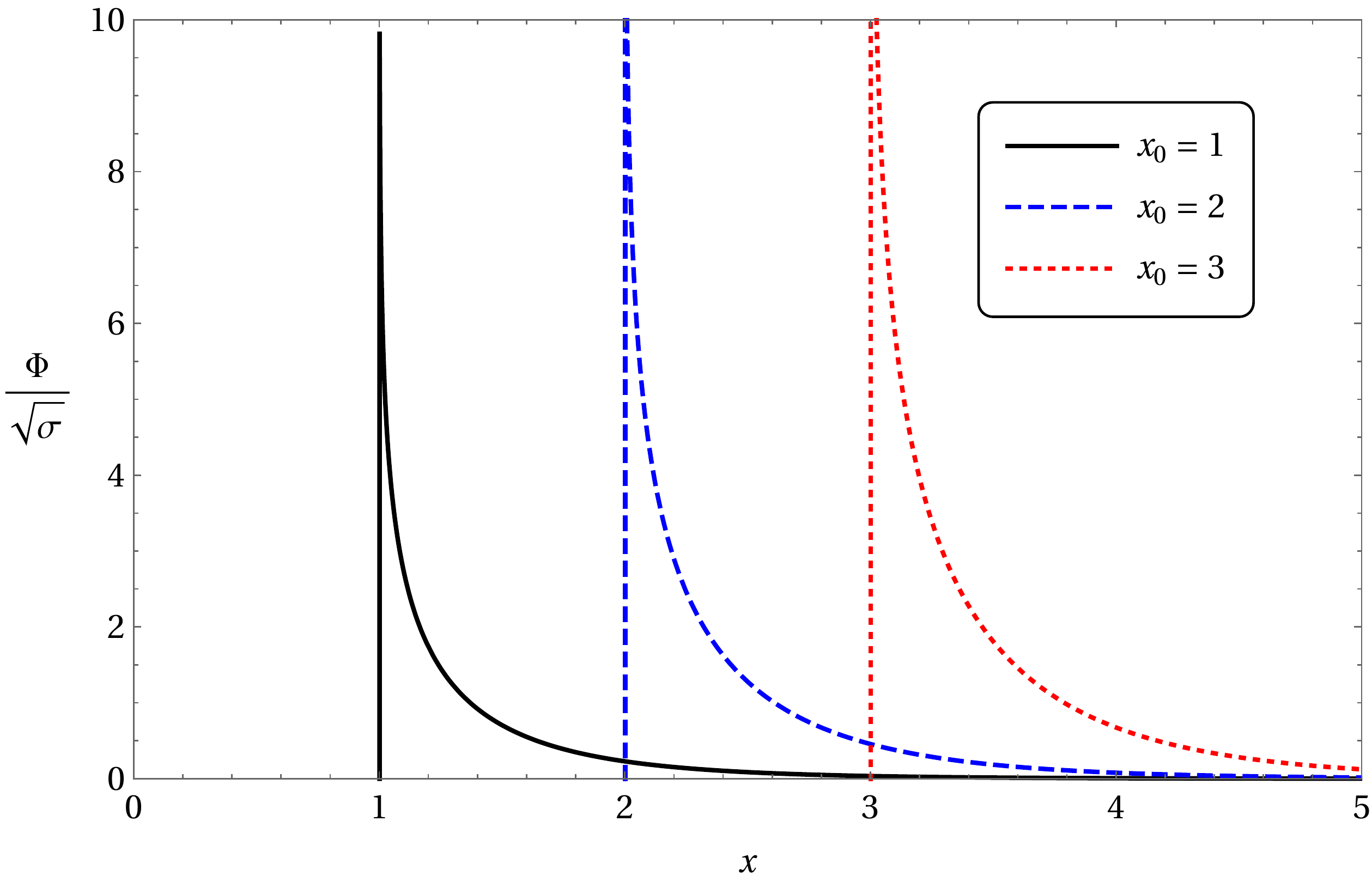}
\caption{Numerical solution for Eq. (\ref{eq:PhiYM1}) for some values of $x_{0}=r_{0}\sqrt{\sigma}$. We assumed $\nu=0.05$, $M_{\text{Cas}}=1.38\sqrt{\sigma}$, and $\Phi_{0}=0$.}\label{PhiDiv}
\end{figure}

Let us turn our attention to the redshift function associated with
the Casimir-Yang-Mills wormhole. By replacing the radial pressure
(\ref{eq:pressureYM}) and the shape function (\ref{eq:brYM}) into
Eq. (\ref{eq:grr}), we find 
\begin{equation}
\Phi(r)=\Phi_{0}+\int_{r_{0}}^{r}\frac{\left(2+\nu+M_{\text{Cas}}\bar{r}\right)\left(M_{\text{Cas}}\bar{r}\right){}^{-\nu-1}e^{-M_{\text{Cas}}\bar{r}}}{2\bar{r}\left[\Gamma\left(-\nu-1,M_{\text{Cas}}\bar{r}\right)-\Gamma\left(-\nu-1,M_{\text{Cas}}r_{0}\right)\right]}d\bar{r},\label{eq:PhiYM1}
\end{equation}
in which the integration is not possible to solve algebraically. Therefore,
we performed numerical integration for different values of throat radius
with $x=r\sqrt{\sigma}$, as shown in Fig. \ref{PhiDiv}. As we can see, the redshift
function is divergent on the throat spoiling the traversable condition
$(iii)$, a situation analogous to the previous case study in section
\ref{sec2}.

\begin{figure}[!h]
\begin{center}
\includegraphics[height=7cm]{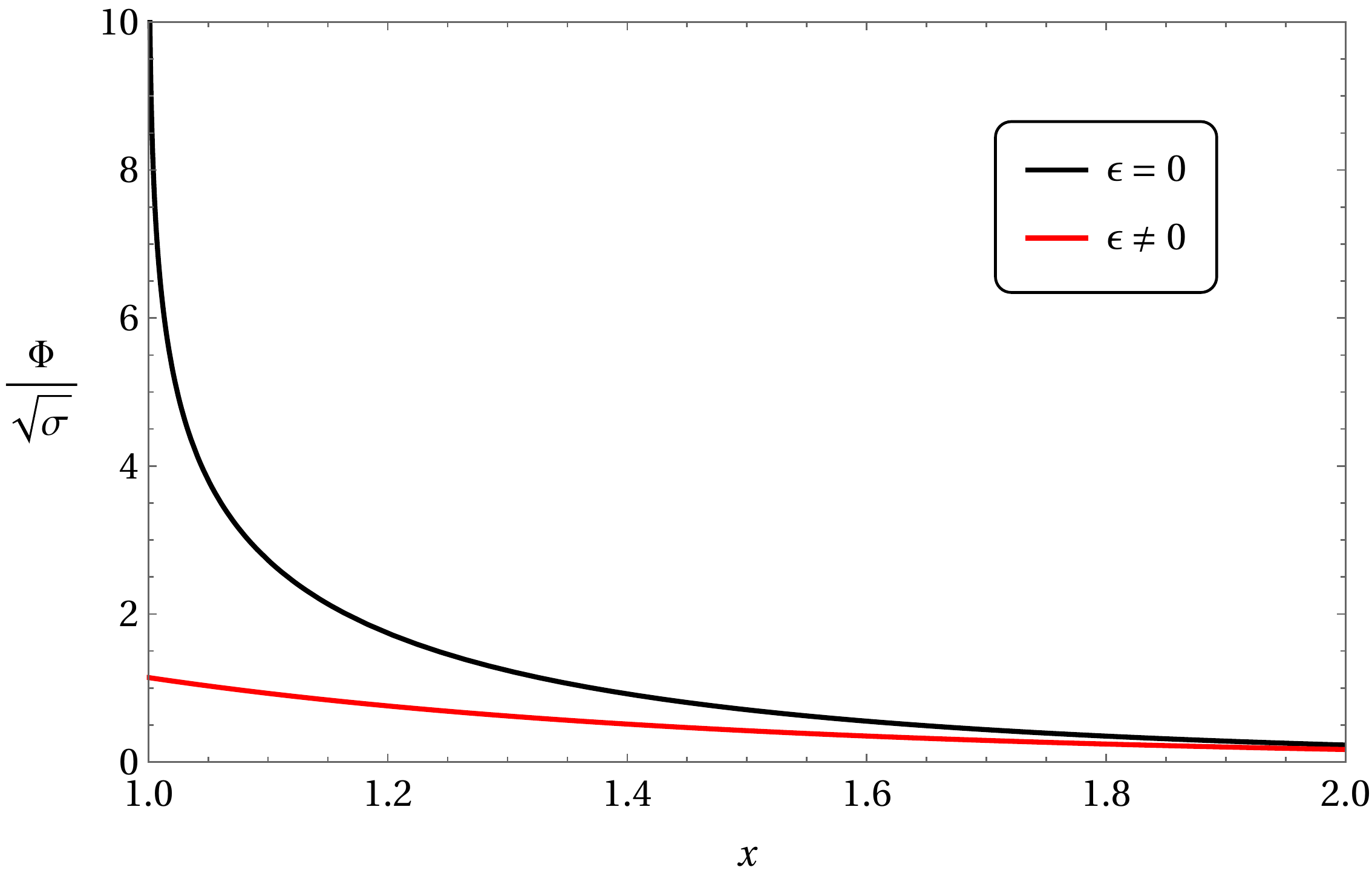}
\caption{Numerical solutions for Eq. (\ref{eq:PhiYM1}) by considering $\epsilon=0$ and $\epsilon=\frac{r_{0}}{M_{\text{Cas}}}\left(2+\nu+M_{\text{Cas}}r_{0}\right)$ for $x_{0}=1$. Our calculations assumed $\nu=0.05$, $M_{\text{Cas}}=1.38\sqrt{\sigma}$, and $\Phi_{0}=0$.\label{figPhifinite}}
\end{center}
\end{figure}

In order to find a traversable wormhole solution, we can apply our deformation method to Eq. (\ref{eq:PhiYM1})
in a similar way to the earlier case. We replaced the EoS parameter
$\omega(r)$ in (\ref{eq:pressureYM}) by
\begin{equation}
\omega(r)\rightarrow\omega(r)-\epsilon\frac{M_{\text{Cas}}}{r^{n}},\label{omegarule2}
\end{equation}
and expanded the integrand in (\ref{eq:PhiYM1}) as a power series
of $\nu$ and $M_{\text{Cas}}$, which resulted in 
\begin{equation}
\Phi_{\epsilon}^{(n)}(r)=\Phi_{0}+\int_{r_{0}}^{r}\frac{r_{0}}{2(\bar{r}-r_{0})^{2}}\left[\frac{(\bar{r}-r_{0})}{\bar{r}^{n+1}}\left(\epsilon M_{\text{Cas}}-\bar{r}^{n}(2+3\nu-M_{\text{Cas}}\bar{r})\right)+2(\nu-M_{\text{Cas}}r_{0})\ln\left(\frac{\bar{r}}{r_{0}}\right)\right]d\bar{r},\label{eq:PhiYMintLie}
\end{equation}
where the higher-order terms have been neglected. The integral can
now be computed for any choice of the positive integer $n$. For example,
if we take $n=1$, we obtain the following $\epsilon$-deformed redshift
function:
\begin{align}
\Phi_{\epsilon}^{(1)}(r) & =\Phi_{0}+\frac{1}{2r_{0}}\left[\frac{\epsilon M_{\text{Cas}}r_{0}}{r}+\frac{2r_{0}^{2}\left(M_{\text{Cas}}r_{0}-\nu\right)}{r-r_{0}}\ln\left(\frac{r}{r_{0}}\right)\right.\nonumber \\
 & \ \ +\left(r_{0}\left(2+\nu+2M_{\text{Cas}}r_{0}\right)-\epsilon M_{\text{Cas}}\right)\ln r\nonumber \\
 & \left.\ -\left(r_{0}\left(2+\nu+M_{\text{Cas}}r_{0}\right)-\epsilon M_{\text{Cas}}\right)\ln\left(r-r_{0}\right)\right].\label{eq:Phidef}
\end{align}
Note that the logarithmic coefficients in lines two and three of Eq.
(\ref{eq:Phidef}) differ due to the Casimir mass correction. Nevertheless,
we can eliminate the term $\ln(r-r_{0})$ that diverges at the throat
by fixing the parameter $\epsilon$ as
\begin{equation}
\epsilon=\frac{r_{0}}{M_{\text{Cas}}}\left(2+\nu+M_{\text{Cas}}r_{0}\right).\label{epsilonYM}
\end{equation}
Hence, at the leading order, the linearized redshift function takes
the form
\begin{equation}
\Phi^{(1)}(r)=\frac{r_{0}}{2r}\left(2+\nu+M_{\text{Cas}}r_{0}\right)+\frac{1}{2}M_{\text{Cas}}r_{0}\ln r-\frac{r_{0}\left(\nu-M_{\text{Cas}}r_{0}\right)}{r-r_{0}}\ln\left(\frac{r}{r_{0}}\right),\label{eq:PhiYM2}
\end{equation}
where we set $\Phi_{0}=0$. In proximity of the throat, we find
\begin{equation}
\lim_{r\rightarrow r_{0}}\Phi^{(1)}(r)=1-\frac{\nu}{2}+M_{\text{Cas}}r_{0}\left(\frac{3}{2}+\frac{1}{2}\ln r_{0}\right),
\end{equation}
is a finite quantity. Therefore, we have demonstrated that the Casimir-Yang-Mills
energy indeed generates a traversable wormhole in $(2+1)$-dimensions.
Also, we may note that, in the massless limit, the expression (\ref{eq:PhiYM2})
agrees with our tree-level result, namely, 
\begin{equation}
\lim_{\begin{smallmatrix}\nu\to0\\
M_{\text{Cas}}\rightarrow0
\end{smallmatrix}}\Phi^{(1)}=\frac{r_{0}}{r}=\Phi_{\text{tree}}^{(1)}.
\end{equation}
It is worth noting that in the asymptotic limit as $r\rightarrow\infty$,
one finds that $\Phi^{(1)}(r)$ diverges due to the presence of the
term $\frac{1}{2}r_{0}M_{\text{Cas}}\ln r$ in (\ref{eq:PhiYM2}).
This is a consequence of the absence of the factor $\exp(-M_{\text{Cas}}\bar{r})$
in the linearized integration (\ref{eq:PhiYMintLie}). However, by applying the substitution (\ref{omegarule2}) into Eq. (\ref{eq:PhiYM1}) and fixing the $\epsilon$-parameter as specified in Eq. (\ref{epsilonYM}), we can numerically integrate the resulting equation and observe that the redshift function is finite at the throat radius and vanishes as $r \rightarrow \infty$. This behavior is illustrated in Figure \ref{figPhifinite}. 

\subsection{Source properties and stability analysis}

\begin{figure}[H]
\begin{center}
\begin{tabular}{ccc}
\includegraphics[height=5cm]{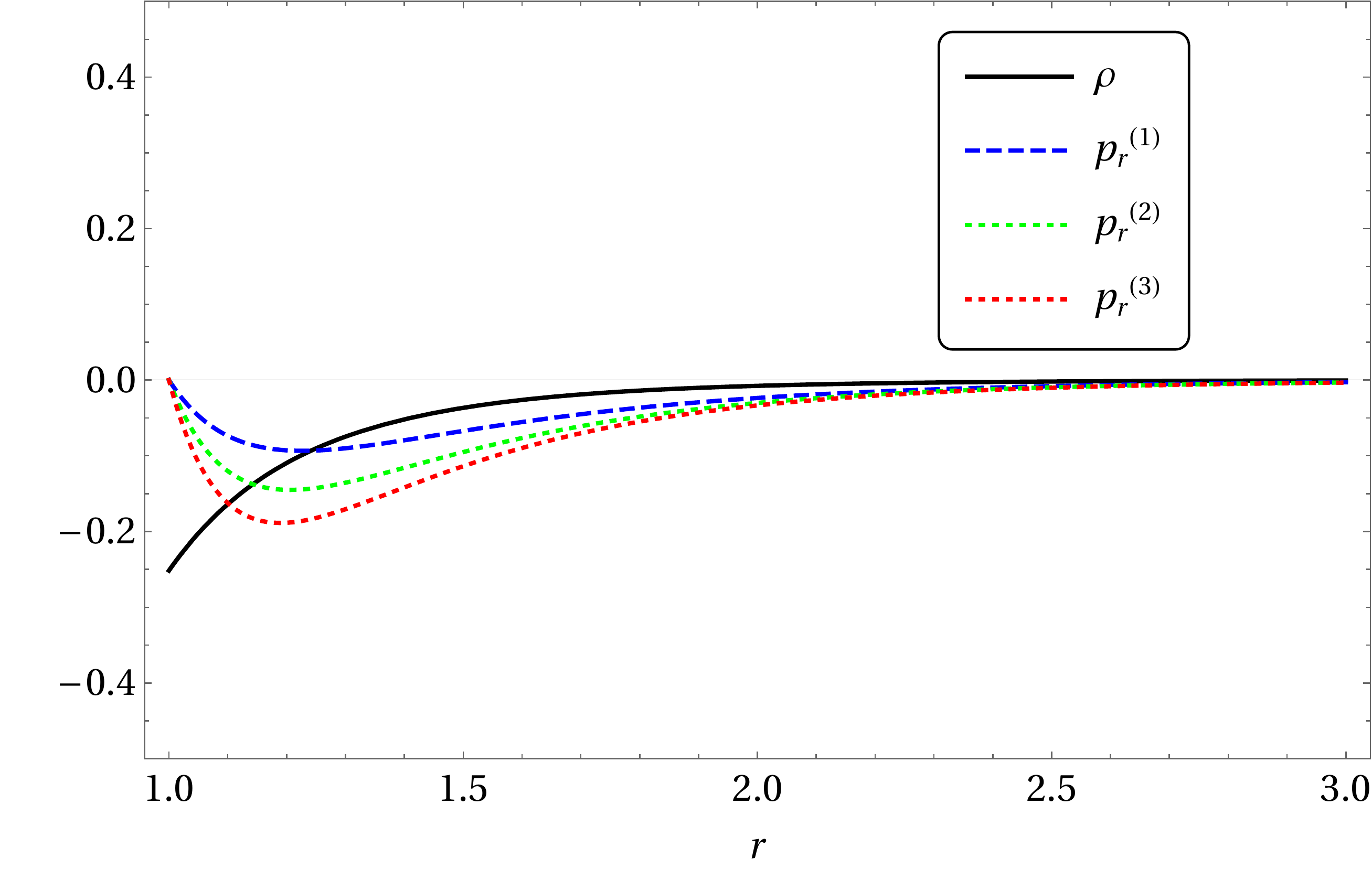}
\includegraphics[height=5cm]{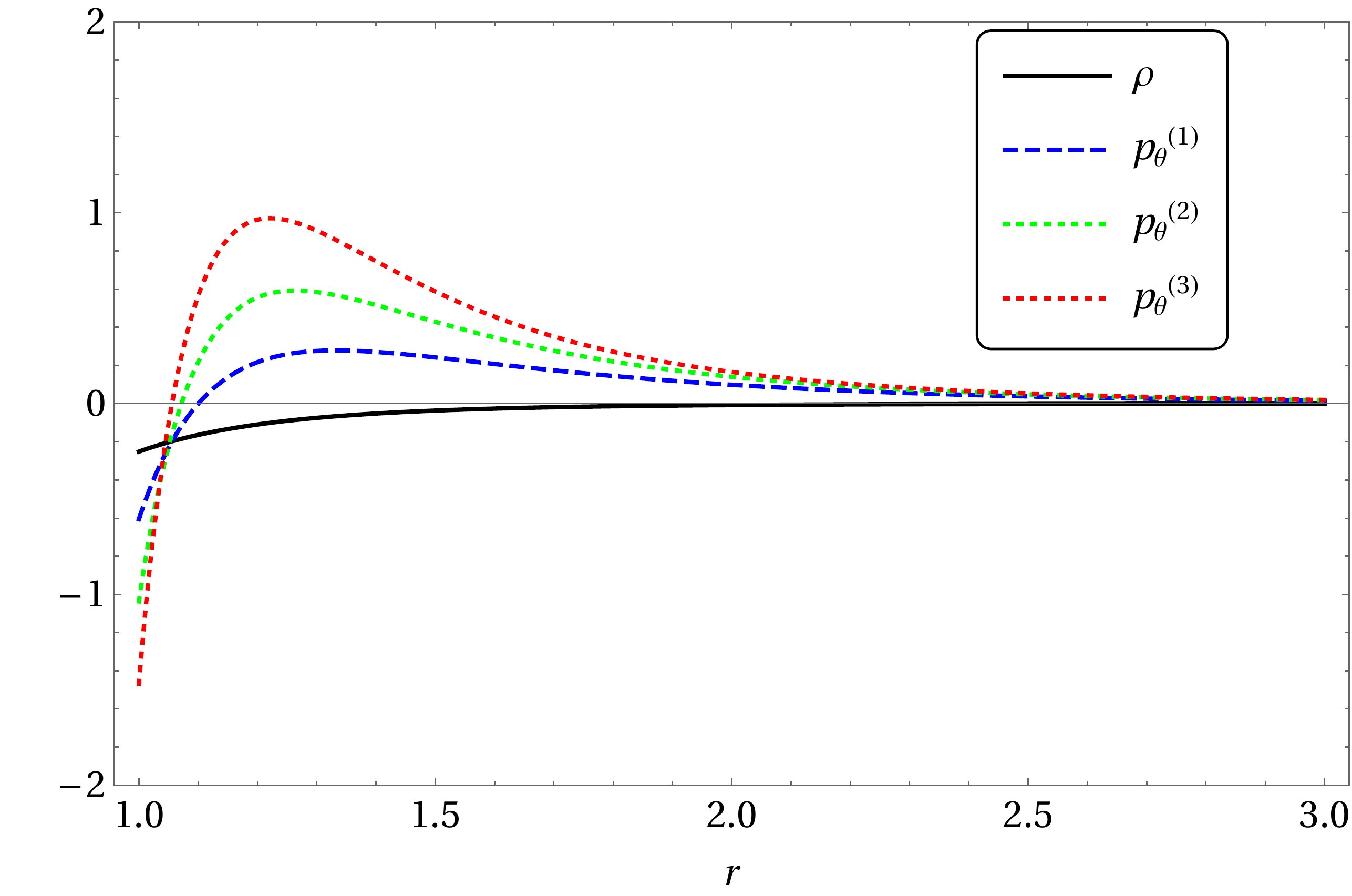}\\ 
(a) \hspace{8 cm}(b)\\
\includegraphics[height=5cm]{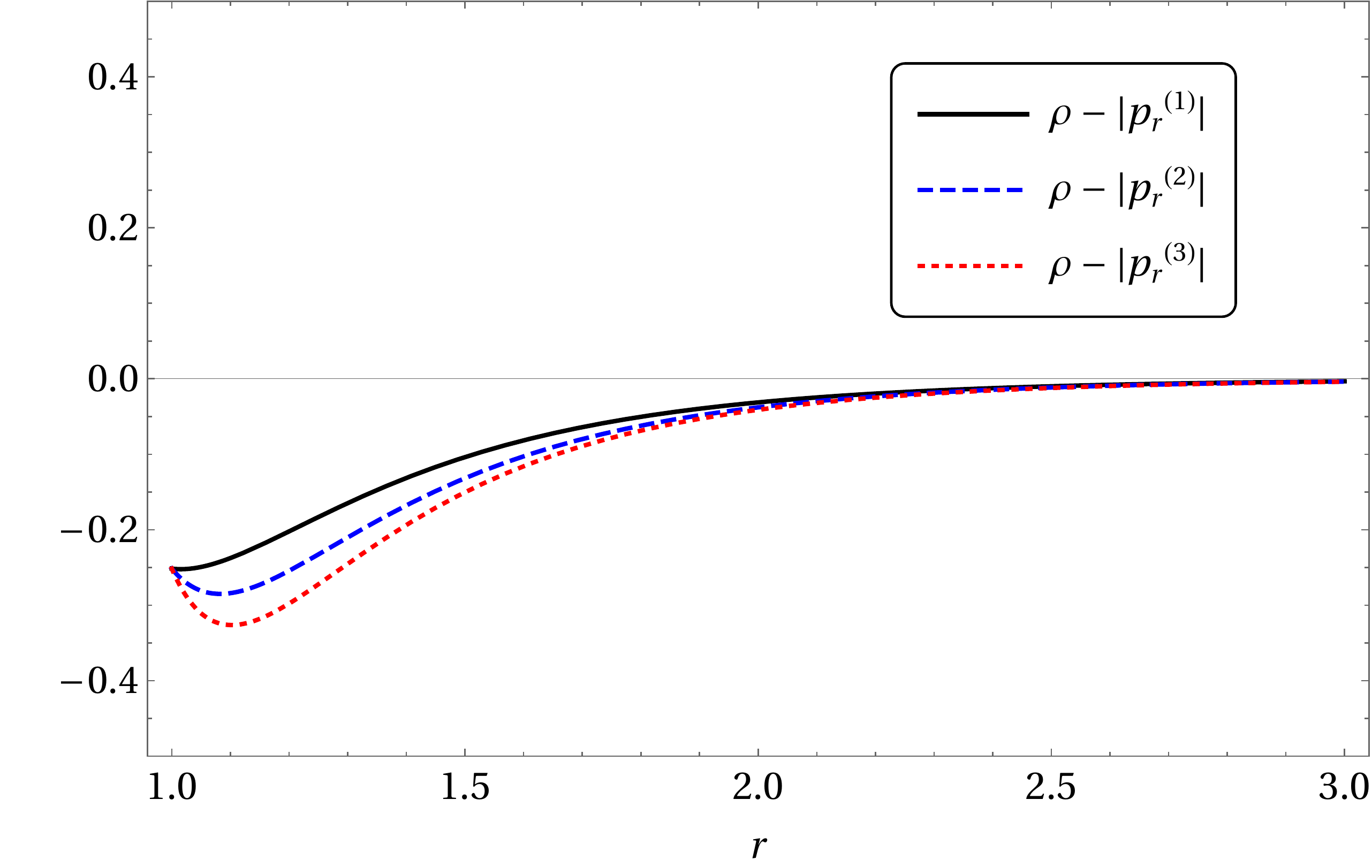}
\includegraphics[height=5cm]{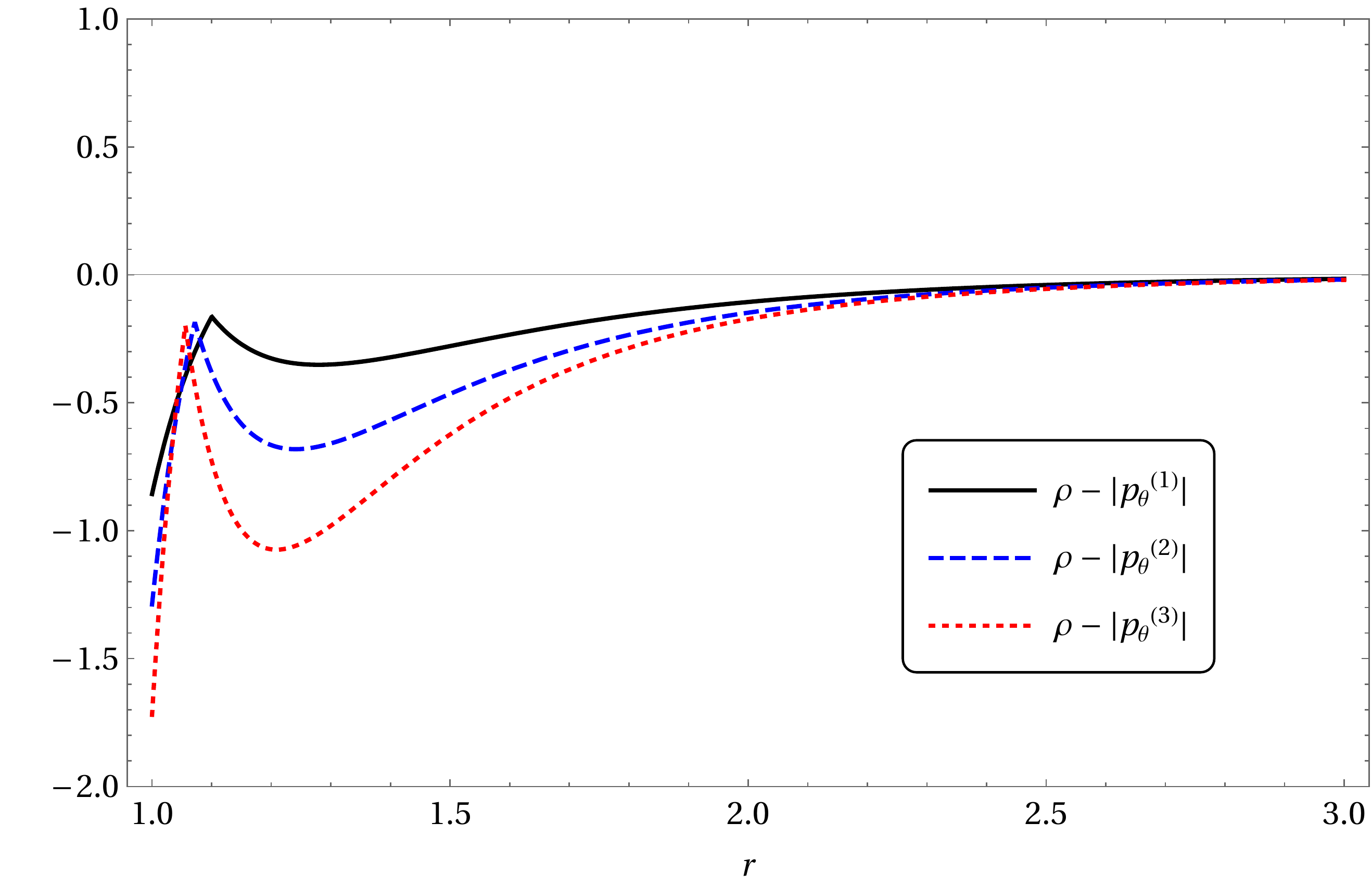}\\ 
(c) \hspace{8 cm}(d)
\end{tabular}
\end{center}
\caption{
Density and radial pressures (a), density and lateral pressures (b), and combination of density and pressures (c) and (d) in order to verify energy conditions, namely, DEC, considering $n=1,2,3$, $\sqrt{\sigma}=1$, $\nu=0.05$, $M_{\text{Cas}}=1.38$, and $r_{0}=\lambda=1$, in Planck units.
\label{fig9}}
\end{figure}

At this point, we can determine all the characteristics associated with the geometry and matter content of these structures. Remembering the shape function of our Casimir-Yang-Mills wormhole solution:
\begin{equation}
b(r)=r+2\kappa\lambda\sigma^{-\frac{\nu}{2}}M_{\text{Cas}}^{\nu+1}r\left[\Gamma(-\nu-1,M_{\text{Cas}}r)-\Gamma(-\nu-1,M_{\text{Cas}}r_{0})\right],\label{eq:brYM2}
\end{equation}and applying our deformation method (\ref{omegarule2}), we found a redshift function to arbitrary $n$ integer, regularized on the throat, given by: 
\begin{equation}
\Phi^{(n)}(r)=\int_{r_{0}}^{r}\frac{M_{\text{Cas}}^{-\nu-1}\bar{r}{}^{-\nu-n-2}\left[\bar{r}^{n}\left(2+\nu+M_{\text{Cas}}\bar{r}\right)-r_{0}^{n}\left(2+\nu+M_{\text{Cas}}r_{0}\right)\right]e^{-M_{\text{Cas}}\bar{r}}}{2\left[\Gamma(-\nu-1,M_{\text{Cas}}\bar{r})-\Gamma(-\nu-1,M_{\text{Cas}}r_{0})\right]}d\bar{r},\label{PhiYMsol}
\end{equation}where we use $\epsilon=\frac{r_{0}^{n}(2+\nu+M_{\text{Cas}}r_{0})}{M_{\text{Cas}}}$. With the energy density coming from the Casimir energy for the $SU(N)$ gauge theory:
\begin{equation}
\rho(r)=-\lambda\frac{\left(r\sqrt{\sigma}\right)^{-\nu}}{r^{3}}e^{-M_{\text{Cas}}r},\label{rhoYM2}
\end{equation}
we derived the radial pressure using the standard relation adopted in the usual Casimir effect:\begin{align}
p_{r}^{(n)}(r) & =\omega_{r}^{(n)}(r)\rho(r),\nonumber\\
 & =\left[2+\nu+M_{\text{Cas}}r-\left(2+\nu+M_{\text{Cas}}r_{0}\right)\left(\frac{r_{0}}{r}\right)^{n}\right]\rho(r).\label{prYM2}
\end{align} Note that, as in the early case, the mean value of the $\omega$-deformation is null:
\begin{equation}
\lim_{r\rightarrow\infty}\frac{1}{r-r_{0}}\int_{r_{0}}^{r}\left[\left(2+\nu+M_{\text{Cas}}r_{0}\right)\left(\frac{r_{0}}{r'}\right)^{n}\right]dr'=0.
\end{equation}And, therefore, on average, again the method maintains the Casimir relations. From Eq. (\ref{eq:gthetatheta}), with the help Eqs. (\ref{eq:brYM2})
and (\ref{PhiYMsol}), the lateral pressure for the Yang-Mills case assumes the form:
\begin{align}
p_{\theta}^{(n)}(r) & =\frac{\lambda\sigma^{-\frac{\nu}{2}}M_{\text{Cas}}^{-\nu-1}r^{-2\nu-2(n+2)}e^{-2M_{\text{Cas}}r}}{2\left[\Gamma\left(-\nu-1,M_{\text{Cas}}r\right)-\Gamma\left(-\nu-1,M_{\text{Cas}}r_{0}\right)\right]}\nonumber \\
 & \times\left\{ \left(rr_{0}\right){}^{n}\left(5+2\nu+2M_{\text{Cas}}r\right)\left(2+\nu+M_{\text{Cas}}r_{0}\right)\right.\nonumber \\
 & \ -r^{2n}\left(2+\nu+M_{\text{Cas}}r\right)\left(3+\nu+M_{\text{Cas}}r\right)\nonumber \\
 & \ -r_{0}^{2n}\left(2+\nu+M_{\text{Cas}}r_{0}\right){}^{2}\nonumber \\
 & \ -2r^{n}r_{0}^{-\nu-1}e^{M_{\text{Cas}}r}\left[r_{0}^{n}\left(2+\nu+M_{\text{Cas}}r_{0}\right)\left(2+n+\nu+M_{\text{Cas}}r\right)\right.\nonumber \\
 & \left.-r^{n}\left(4+4\nu+\nu^{2}+3M_{\text{Cas}}r+2\nu M_{\text{Cas}}r+M_{\text{Cas}}^{2}r^{2}\right)\right]\nonumber \\
 & \ \times\left.\left[r_{0}^{\nu+1}E_{\nu+2}(M_{\text{Cas}}r)-r^{\nu+1}E_{\nu+2}(M_{\text{Cas}}r_{0})\right]\right\} ,\label{Ptheta2}
\end{align} where $E_{n}(z)=\int_{1}^{\infty}\frac{e^{-zt}}{t^{n}}dt$ gives the
exponential integral function. As expected, the EoM (\ref{eq:g00}-\ref{eq:gthetatheta}) and the conservation
law (\ref{eq:conservation}) are satisfied by the above expressions. Finally, we analyzed the energy conditions and observed that, surprisingly, despite having different relations from the noninteracting case, the interacting case exhibits extremely similar graphical behavior, as evidenced in Fig. \ref{fig9}. Once again, we have the violation of the NEC, DEC, and also of the SEC.

\subsubsection{Stability from sound velocity}

\begin{figure}[!]
\begin{center}
\includegraphics[height=8cm]{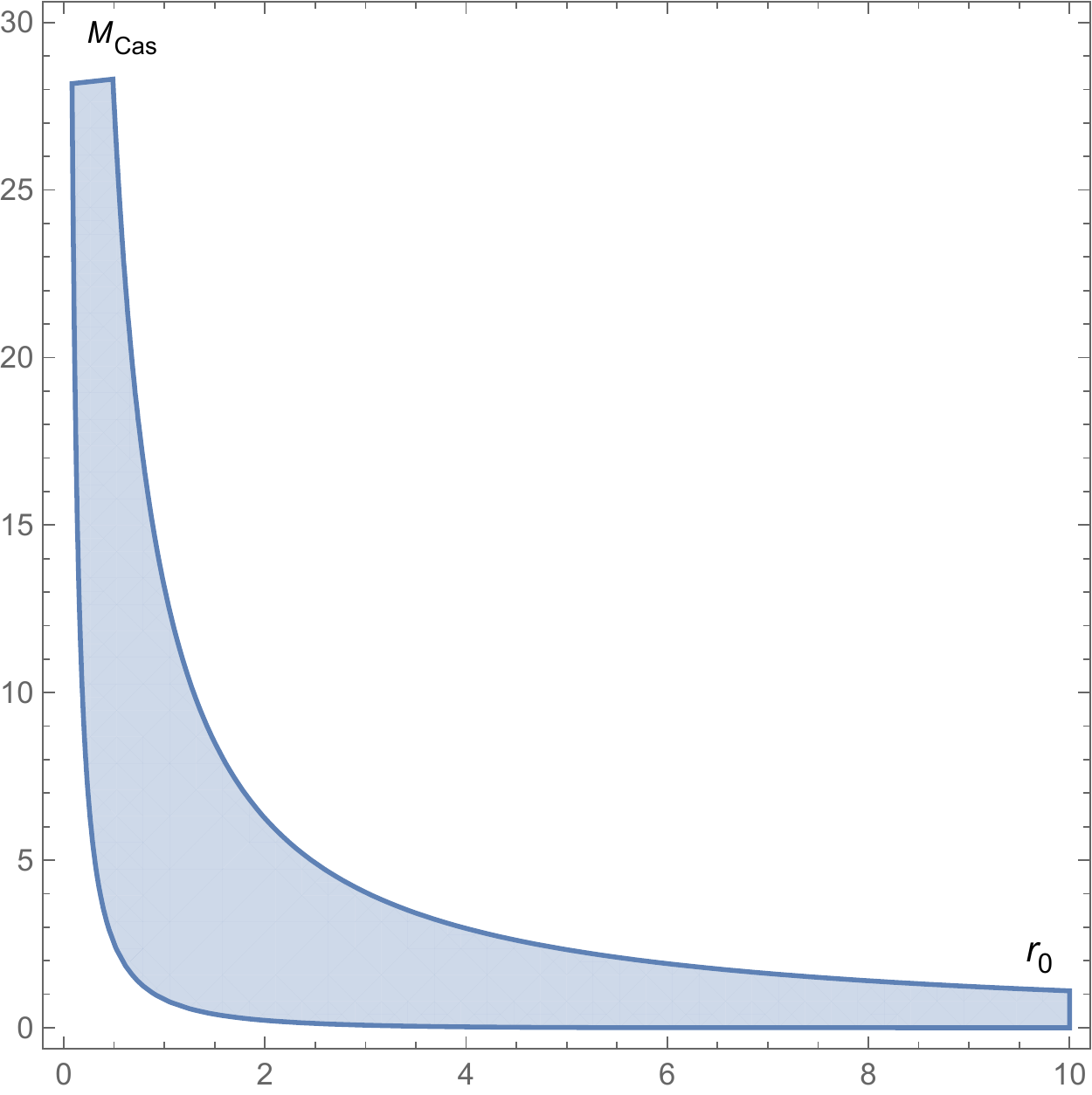}
\caption{Parameter space ($r_0, M_{\text{Cas}}$) nearby the wormhole throat ($r\approx r_0$) for the interacting scenario, highlighting the region in which $0 \leq v_s^2 \leq 1$, with $\nu = 0.05$, $n = 1$, $\lambda = \frac{3 \zeta(3)}{16 \pi}$. The $\sigma$ parameter is constrained by Eq. (\ref{eq:flatness}).}\label{Sound1}
\end{center}
\end{figure}

The stability analysis of our solutions, based on the examination of sound velocity within the source as defined by Eq. (\ref{Sound}), demonstrates a consistent alignment with the noninteracting case previously studied. Specifically, we observe sustained stability near the wormhole throat, albeit with a diminishing stability region as the power parameter $n$ increases. This suggests an increasing susceptibility to instability as the solution approaches the exact Casimir configuration, in line with the noninteracting scenario. However, a departure from this latter is found in instances where the sound velocity falls below the speed of light for certain parameter values, a phenomenon absent in the noninteracting case. The visual representation of this behavior is provided in Figure \ref{Sound1}, illustrating the parameter space region where $0 \leq v_s^2 \leq 1$, denoting stability and physical validity.

\subsubsection{Stability from TOV equation}

\begin{figure}[t!h]
\begin{center}
\begin{tabular}{ccc}
\includegraphics[height=5.1cm]{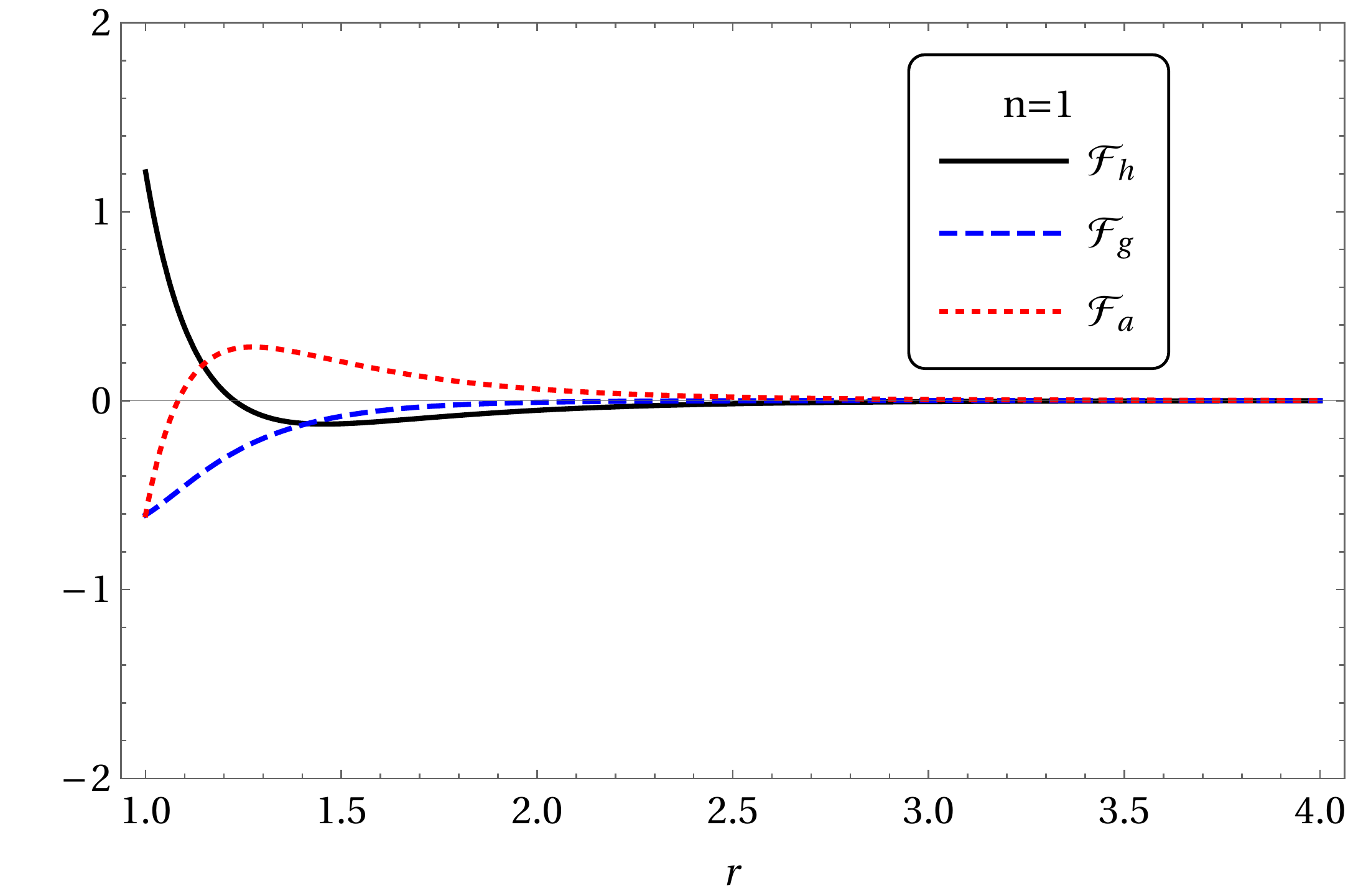}
\includegraphics[height=5cm]{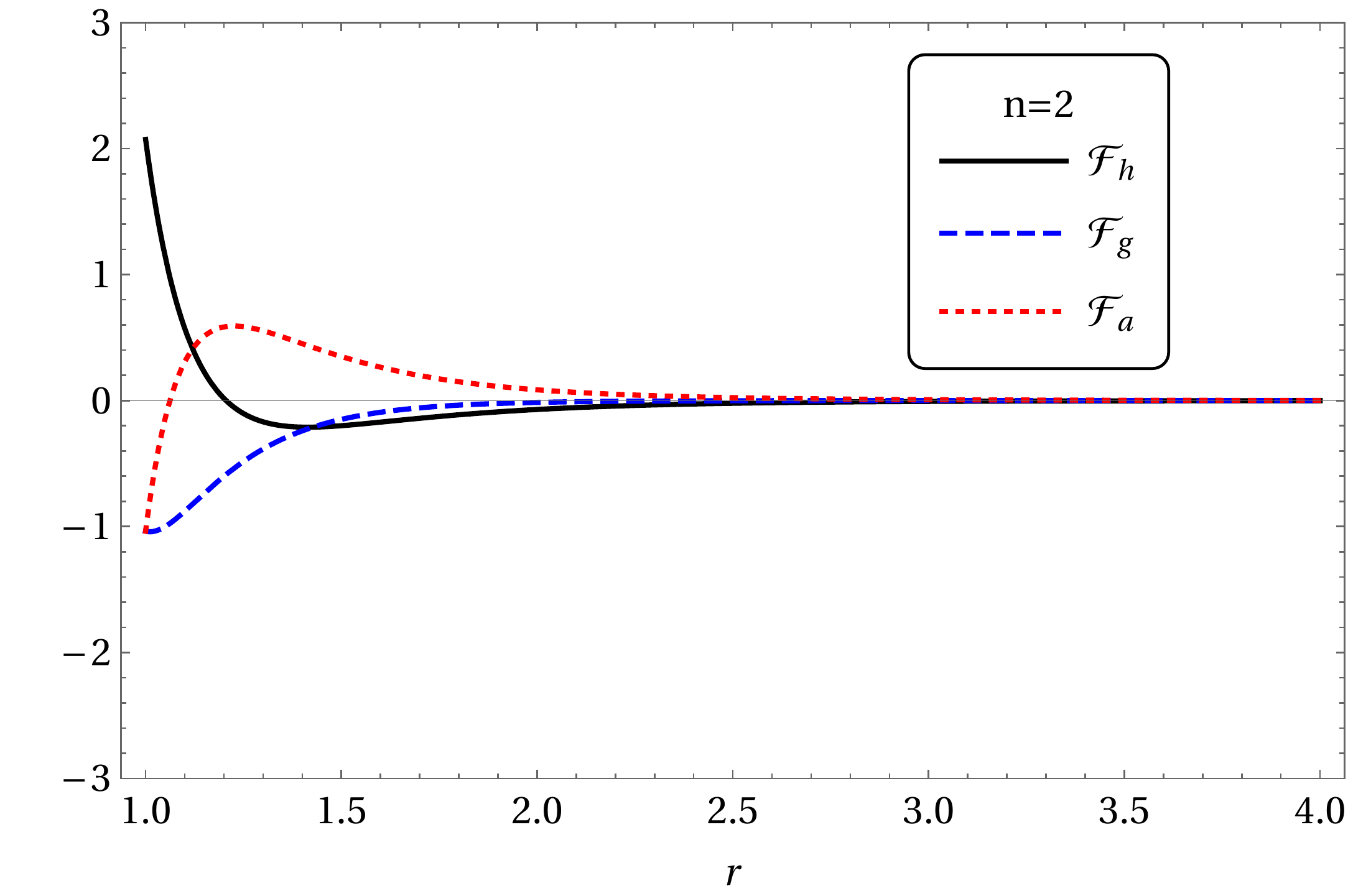}\\ 
(a) \hspace{8 cm}(b)
\end{tabular}
\end{center}
\caption{
The graphical representation of $\mathcal{F}_{h}$, $\mathcal{F}_{g}$, and $\mathcal{F}_{a}$ as functions of the radial coordinate $r$, with  $\sqrt{\sigma}=1$, $\nu=0.05$, $M_{\text{Cas}}=1.38$, and $r_{0}=\lambda=1$, in Planck units.\label{figForceYM}}
\end{figure}

Similar to the previous case, the stability analysis for the non-abelian version of the Casimir wormhole can be conducted with the help of the TOV equation (\ref{TOV1}). By utilizing Eqs (\ref{PhiYMsol}), (\ref{rhoYM2}), (\ref{prYM2}) and (\ref{Ptheta2}), we can precisely determine the forces acting on the matter comprising our wormhole solution for any positive integer value of $n$, specifically,
\begin{eqnarray}
\mathcal{F}_{h}&=&-\frac{\lambda e^{-M_{\text{Cas}}r}}{\sigma^{\nu/2}r^{4+\nu}}\nonumber\\
	&&\times\left[6+\nu^{2}+\left(5+2M_{\text{Cas}}r\right)\nu+\left(4+M_{\text{Cas}}r\right)M_{\text{Cas}}r\right.\nonumber\\
	&&\left.-\left(\frac{r_{0}}{r}\right)^{n}\left(2+\nu+M_{\text{Cas}}r_{0}\right)\left(3+n+\nu+rM_{\text{Cas}}\right)\right],
\end{eqnarray}
\begin{eqnarray}
\mathcal{F}_{g} &=& \frac{\lambda\left(\sqrt{\sigma}M_{\text{Cas}}\right)^{-\nu}\left[r^{n}\left(2+\nu+M_{\text{Cas}}r\right)-r_{0}^{n}\left(2+\nu+M_{\text{Cas}}r_{0}\right)\right]e^{-2M_{\text{Cas}}r}}{2M_{\text{Cas}}r^{5+n+2\nu}\left[\Gamma\left(-\nu-1,M_{\text{Cas}}r\right)-\Gamma\left(-\nu-1,M_{\text{Cas}}r_{0}\right)\right]}\nonumber\\
&&\times \left[3+\nu+M_{\text{Cas}}r-\left(2+\nu+M_{\text{Cas}}r_{0}\right)\left(\frac{r_{0}}{r}\right)^{n}\right],
\end{eqnarray}and
\begin{eqnarray}
\mathcal{F}_{a}	&=&\frac{1}{2}\lambda\sigma^{-\frac{\nu}{2}}r^{-5-2(n+\nu)}e^{-2M_{\text{Cas}}r}\nonumber\\
&&\times	\left\{ 2e^{M_{\text{Cas}}r}r^{1+n+\nu}\left[r^{n}\left(4+\nu^{2}+2\nu\left(2+M_{\text{Cas}}r\right)+M_{\text{Cas}}r\left(3+rM_{\text{Cas}}\right)\right)\right.\right.\nonumber\\
	&&\left.-r_{0}^{n}\left(2+\nu+M_{\text{Cas}}r_{0}\right)\left(2+n+\nu+M_{\text{Cas}}r\right)\right]\nonumber\\
	&&+\frac{M_{\text{Cas}}^{-1-\nu}\left(r^{n}\left(2+\nu+M_{\text{Cas}}r\right)-r_{0}^{n}\left(2+\nu+M_{\text{Cas}}r_{0}\right)\right)}{\Gamma\left(-\nu-1, M_{\text{Cas}r}\right)-\Gamma\left(-\nu-1,M_{\text{Cas}}r_{0}\right)}\nonumber\\
	&&\times\left[r_{0}^{n}\left(2+\nu+M_{\text{Cas}}r_{0}\right)+2r^{n}e^{M_{\text{Cas}}r}E_{2+\nu}\left(M_{\text{Cas}}r\right)\right.\nonumber\\
	&&\left.\left.-r^{n}\left(3+\nu+M_{\text{Cas}}r+2e^{M_{\text{Cas}}r}E_{2+\nu}\left(M_{\text{Cas}}r_{0}\right)\left(\frac{r}{r_{0}}\right)^{1+\nu}\right)\right]\right\} .
\end{eqnarray}Once again, we have verified that the equilibrium condition (\ref{TOV2}) is satisfied for all $r$, assuming that the parameters of the model, such as $\lambda$, $\sigma$, $\nu$, and $M_{\text{Cas}}$, are real and positive quantities (using the computer algebra program \cite{Martin-Garcia:2008yei}). Under this criterion, our solution represents a stable wormhole. 
Figure \ref{figForceYM} demonstrates that the equilibrium stage can be achieved due to the combined effect of pressure hydrostatic, gravitational, and anisotropy forces. It is worth noting that the value of $\mathcal{F}_{h}$ is positive, while $\mathcal{F}_{g}$ and $\mathcal{F}_{a}$ are negative near the throat, resulting in a balanced system when they are combined, resembling the behavior in the noninteracting case.

\section{Conclusion \label{conclusion}}

In summary, our study has led to the discovery of three-dimensional traversable wormhole solutions sourced by the Casimir density and pressure related to the quantum vacuum fluctuations of the Yang-Mills field. We have initiated the analysis at the tree level (from the noninteracting Casimir energy) with an arbitrary state parameter of $\omega$. The determination of the throat radius through the requirement of asymptotic flatness is an important feature of these solutions, and we have found that it is on the order of $r_{0}\sim\tilde{G}$ (in natural units). The constant shape function equal to this radius further highlights the elegance of those solutions.

Our novel methodology of deforming the state parameter $\omega$ by adding a small function obeying an inverse power law of the radial coordinate has allowed us to find finite redshift functions throughout all space. This methodology is critical to understanding the wormhole's behavior since, without it, the object behaves as a black hole for $\omega<0$ and is not a wormhole for other values of this parameter. Additionally, our analysis reveals that the space-averaged deformation function approaches zero, indicating that the object behaves as a legitimate Casimir wormhole with an expected average state parameter of $\omega=2$. This behavior becomes more pronounced as the power parameter $n$ increases.

Our study has also revealed valuable insights into the curvature properties of the obtained family of wormhole solutions. Specifically, we have found that the Kretschmann scalar reaches a global maximum at the wormhole's throat and increases with the power parameter $n$. We have also observed a local maximum in the Kretschmann scalar that increases with $n$. Although the local minimum of the Kretschmann scalar shifts slightly towards the throat with increasing $n$, it remains on the curve corresponding to $n=1$.

We have also examined the null energy condition (NEC) and the dominant energy condition (DEC) and found that these conditions are violated throughout all space. The strong energy condition (SEC) becomes not satisfied since the NEC is broken. These results emphasize the need for further investigation into the stability and viability of these solutions in physical scenarios.

Next, we examined the stability of this specific family of Casimir wormholes, focusing on the squared sound velocity in the surrounding medium. The results demonstrated that the obtained traversable solutions are stable near the wormhole throat, and the size of the stable region diminishes as the parameter $n$ increases. However, these noninteracting solutions were found to be unphysical since the sound speed consistently exceeded the speed of light near the wormhole throat, irrespective of the selected parameter values. In the context of the TOV equation, our analysis confirmed the system's stability across all spatial domains.

We also have discovered a novel family of traversable wormhole solutions, utilizing the quantum vacuum fluctuations of interacting Yang-Mills fields as a source. Through analytical derivation of the shape function, we have confirmed its compliance with all the requirements for generating a traversable wormhole. We have successfully obtained a well-behaved redshift function by modifying the effective state parameter, following a similar approach as in our previous examination. These wormhole solutions exhibit sensitivity to system parameters such as the radius of the wormhole throat, Casimir mass, anomalous exponent, and string tension. Notably, our research has revealed that higher string tension results in a larger throat radius, which can be attributed to the stretching effect in an attempt to deconfine the gluons.

In the sequel, we also investigated the stability of interacting Y-M Casimir wormholes based on squared sound velocity within the source. We found stable solutions near the wormhole throat, with diminishing stability as they approached the Casimir configuration (for increasing $n$ powers), mirroring the noninteracting case. Notably, certain parameter values led to sound velocity dropping below the speed of light, differentiating these solutions and enhancing their physical relevance. Thus, in the context of the non-interacting scenario, subluminal velocities are observed alongside stability only within a limited region around the wormhole throat. In this case, on considering the other space regions, as the power parameter $n$ increases, superluminal velocities, and instability become more pronounced, similar to the increase in energy condition violations, as we can compare Fig. \ref{fig2} with Fig. \ref{Sound2}. On the other hand, in the interacting scenario, stability and subluminal velocities coexist across the entire space, depending on specific parameter choices. Notably, this coexistence persists irrespective of energy condition violations, as previously demonstrated. Moreover, considering the TOV equation revealed consistent stability across all space, akin to the prior case.

In conclusion, our results reveal new possibilities for constructing Casimir wormhole solutions and contribute to our understanding of their properties.

\section*{Acknowledgments}
\hspace{0.5cm} The authors thank the Funda\c{c}\~{a}o Cearense de Apoio ao Desenvolvimento Cient\'{i}fico e Tecnol\'{o}gico (FUNCAP), the Coordena\c{c}\~{a}o de Aperfei\c{c}oamento de Pessoal de N\'{i}vel Superior (CAPES), and the Conselho Nacional de Desenvolvimento Cient\'{i}fico e Tecnol\'{o}gico (CNPq), Grants no. 88887.822058/2023-00 (ACLS), no. 308268/2021-6 (CRM), and no. 200879/2022-7 (RVM) for financial support. R. V. Maluf thanks the Department of Theoretical Physics $\&$ IFIC of the University of Valencia - CSIC for the kind hospitality. The authors also thank the anonymous referee for his valuable comments and suggestions.


\end{document}